\documentclass[aps,prd,reprint,showpacs,showkeys,floatfix]{revtex4-1}
\pdfoutput=1
\usepackage{amssymb,amsmath,graphics,epsfig}

\newcommand{\lowres}{}
\renewcommand{\lowres}{-}

\newcommand{\rem}[1]{ }
\newcommand{\beq}{\begin{equation}}
\newcommand{\eeq}{\end{equation}}
\newcommand{\bea}{\begin{eqnarray}}
\newcommand{\eea}{\end{eqnarray}}

\newcommand{\bra}[1]{\left< #1\right|}
\newcommand{\ket}[1]{\left| #1\right>}

\def\la{{\lesssim}}

\def\apj{{\it Astrophys. J.}}

\def\prl{{\it Phys. Rev. Lett.}}

\def\physrep{{\it Phys. Rep.}}

\def\jpha{{\it J. Phys. A: Math. General}}
\def\aa{{\it Adv. in Astron.}}

\def\jetp{{\it Sov. J. Exp. Theor. Phys.}}
\def\jetpl{{\it Sov. J. Exp. Theor. Phys. Lett.}}
\def\ajp{{\it Am. J. Phys.}}
\def\arnps{{\it Ann. Rev. Nucl. Part. Sci.}}
\def\mpl{{\it Mod. Phys. Lett.}}
\def\physatomnucl{{\it Phys. Atom. Nucl.}}

\begin{document}

\title{On the Dynamics of Non-Relativistic Flavor-Mixed Particles}
\author{Mikhail V. \surname{Medvedev}$^{1,2,3}$} 
\email{mmedvedev@cfa.harvard.edu}
\affiliation{$^1$Department of Physics and Astronomy, University of Kansas, Lawrence, KS 66045}
\affiliation{$^2$Institute for Theory and Computation, Harvard University, 60 Garden St., Cambridge, MA 02138}
\affiliation{$^3$ITP, NRC ``Kurchatov Institute", Moscow 123182, Russia}

\begin{abstract}
Evolution of a system of interacting non-relativistic quantum flavor-mixed particles is considered both theoretically and numerically. It was shown that collisions of mixed particles not only scatter them elastically, but can also change their mass eigenstates thus affecting particles' flavor composition and kinetic energy. The mass eigenstate conversions and elastic scattering are related but different processes, hence the conversion $S$-matrix elements can be arbitrarily large even when the elastic scattering $S$-matrix elements vanish. The conversions are efficient when the mass eigenstates are well-separated in space but suppressed if their wave-packets overlap; the suppression is most severe for mass-degenerate eigenstates in flat space-time. The mass eigenstate conversions can lead to an interesting process, called `quantum evaporation,' in which mixed particles, initially confined deep inside a gravitational potential well and scattering only off each other, can escape from it without extra energy supply leaving nothing behind inside the potential at $t\to \infty$. Implications for the cosmic neutrino background and the two-component dark matter model are discussed and a prediction for the direct detection dark matter experiments is made.  
\end{abstract}
\maketitle

\section{Introduction}

A number of known and hypothetic particles are flavor-mixed, e.g., neutrinos, kaons, quarks, a neutralino, an axion (can be mixed with a photon), to name a few. How these particles behave in the non-relativistic limit has not been carefully studied for years. This paper addresses some important aspects of this profound physical problem.

Mass (propagation) and flavor (interaction) eigenstates are the vectors obtained by diagonalizing the propagation and interaction parts of particle's Hamiltonian, respectively, and they can generally be not identical but related through a unitary transformation 
\beq
\ket{f_i}=\sum_j U_{ij}\ket{m_j}, 
\eeq
where $\ket{f}$ and $\ket{m}$ denote the flavor and mass eigenstates, and $U$ is a unitary matrix. Hence, a mixed particle produced in a reaction has a specific flavor eigenstate, $\alpha$, described by a wave-function being a superposition of several mass eigenstates  \cite{Pontecorvo58}. When a mixed particle is propagating, the mass eigenstates move with different velocities, which causes time-dependent interference known as flavor oscillations. 

An interesting and rather counter-intuitive property of non-relativistic flavor-mixed particles has been found, which is illustrated in the following example. Let us create a non-relativistic electron neutrino in a gravitational potential well. One should expect that if the neutrino is initially confined in the potential, it will remain confined forever (flavor oscillations do not change the picture). However, this is not so if the neutrino scatters elastically off other non-mixed particles from time to time. It has been shown that there is a non-vanishing probability to detect this electron neutrino outside the potential at a later time, although no extra energy has been supplied to it \cite{M10}.  This effect, referred to as the ``quantum evaporation'', is associated with mass eigenstate conversions --- another process discussed in Ref. \cite{M10} --- which we will often refer to as the ``$m$-conversion'' or ``$m$-process'', for brevity. In our example here, a conversion of a heavier mass eigenstate yields a lighter one with a larger velocity. If this velocity exceeds the escape velocity, the light mass eigenstate is unbound and escapes to infinity. Of course, the time scale for scattering has to be less than that for the eigenstate separation to allow this cycle to proceed. Note, however, that evaporation in such a thought experiment is not complete: only the heavy eigenstate can be converted into the escaping lighter eigenstate, whereas the initially created least massive eigenstate remains always bound if it was bound initially. We underscore that such quantum evaporation and $m$-conversions, proposed in \citep{M10} and further elaborated here, have no relation to vacuum flavor oscillations or oscillations in matter whatsoever.

In this paper, we explore elementary processes in an ensemble of flavor-mixed particles. Specifically, we consider a system of two non-relativistic flavor-mixed particles confined inside the gravitational potential well which can scatter off each other. We demonstrate that complete evaporation of both these particles is possible in this case. Indeed, when a bound mixed particle scatters off normal matter, only the heavy eigenstate can be $m$-converted with the increase of its kinetic energy ({\it  a l\'a} an exothermic reaction) and ultimately escape. Conversions of the least heavy eigenstates are always ``endothermic", hence the trapped ones will never get enough speed to escape. However, if conversions occur in interaction of two mass eigenstates of two flavor-mixed particles, then the trapped lightest state can get substantial {\em recoil} velocity and escape. It is this process that opens up a possibility of a {\em complete} evaporation of an ensemble of mass-eigenstates. We also show in the paper that the scattering and conversion transition amplitudes are fairly independent, so it is possible to have conversions even when the scattering $S$-matrix elements vanish. Finally, we demonstrate that the $m$-conversion amplitude (and hence its cross-section) in Minkowsky space (without gravity) is strongly suppressed in the mass-degenerate case. These results are important for better understanding of the properties of mixed particle in general, as well as have interesting implications for the cosmic neutrino background and, possibly, dark matter physics and cosmology.

\section{Interacting mixed particles}

In this paper we are interested in interactions of individual mass eigenstates. A mixed particle is created in a flavor state, but it consists of several mass eigenstates. Although all of these `pieces' comprise a single particle, it is possible to visualize their kinematics as if they were normal particles having different masses. In general, these mass eigenstates propagate with different velocities and in the non-relativistic limit they separate from each other rapidly. (This is quite different form the relativistic case in which all eigenstates propagate nearly at the speed of light and it may take a while for them to separate, hence the plane wave approximation is commonly used.) Therefore, fairly soon, a mixed particle becomes a collection of spatially separated mass eigenstates which can interact with other particles independently. Apparently, an ensemble of non-relativistic flavor-mixed particles is, in most cases, an ensemble of individual mass eigenstates. Therefore, it is very natural to investigate the evolution of such an ensemble in the mass basis rather than the flavor basis, which is usually used. The interaction matrix, however, is non-diagonal in the mass basis and off-diagonal terms represent transitions between different mass eigenstates. Should the mass eigenstates overlap to represent a particular flavor, these off-diagonal couplings `balance' transitions of mass eigenstates into each other precisely to produce the scattered particle in a flavor eigenstate again. In contrast, if an individual mass eigenstate interacts, there is no such a `balance', so new (absent) mass eigenstates are produced. Thus, one mass eigenstate can be converted into others. Such a process of a $m$-conversion is of primary interest to us.

\begin{figure}
\includegraphics[width=85mm]{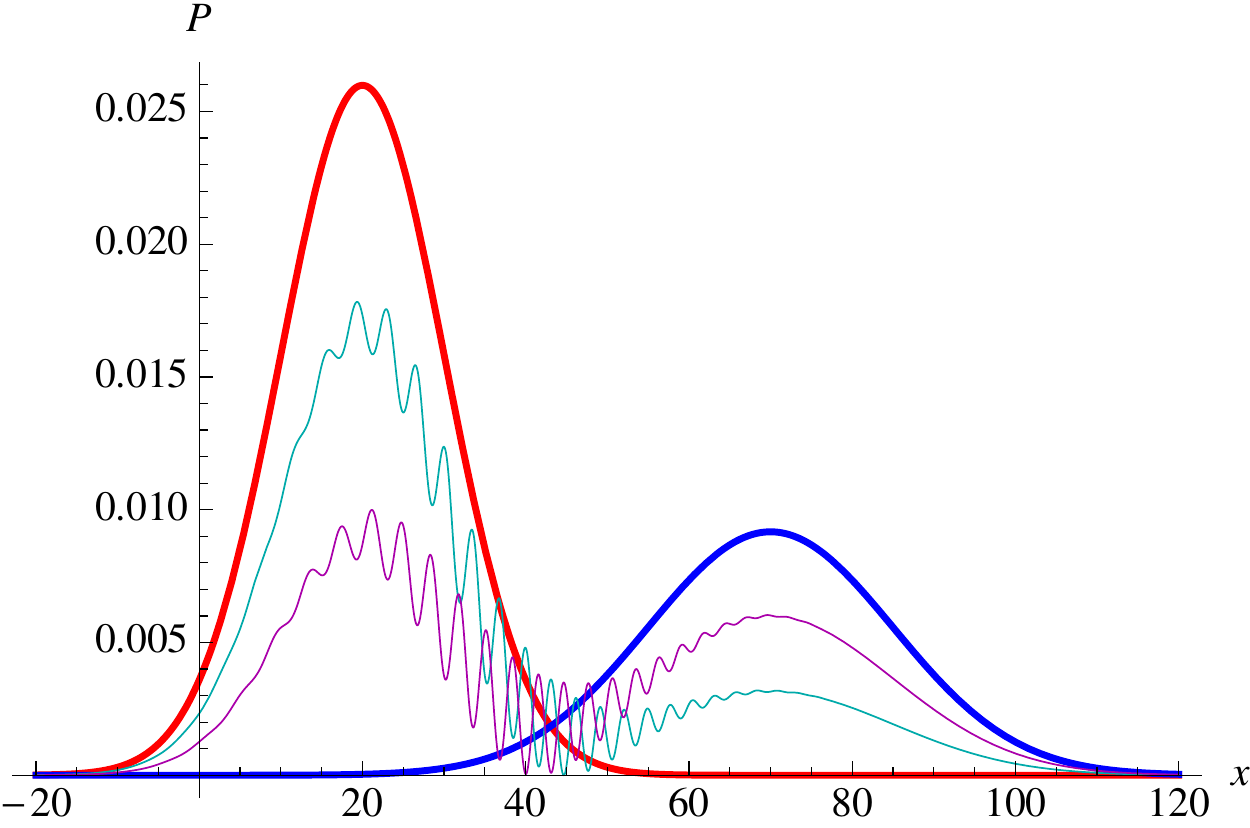}
\caption{Probability distributions of different eigenstates in space. The bold curves (red and blue) represent heavy and light mass eigenstates and thin curves (cyan and magenta) are flavor eigenstates. Flavor oscillations are seen in the overlap region.
\label{wavepackets}}
\end{figure}

Here we consider a simple model of stable two-flavor particles. We will interchangeably denote flavor eigenstates as $\ket{f_\alpha}$ and $\ket{f_\beta}$ or as just $\alpha$ and $\beta$, whenever it's not confusing. Since masses of the mass eigenstates are different, we refer to them as heavy and light eigenstates, hence $m_l<m_h$. Thus, similarly, the mass eigenstates are denoted as $\ket{m_h}$ and $\ket{m_l}$ or as just $h$ and $l$. A two-component flavor-mixed particle is described by a two-component wave-function, which representations in the flavor and mass bases are related via a $2\times2$ rotation matrix, $U$, where $\theta$ is the mixing angle, i.e.,
\beq
{\alpha \choose \beta}=
\left(\begin{matrix}
\cos\theta & \sin\theta \\
-\sin\theta & \cos\theta
\end{matrix}\right)
{h \choose l}.
\label{mix}
\eeq
Fig. \ref{wavepackets} illustrates such a particle. The bold red and blue curves represent heavy and light mass eigenstates assumed to have gaussian wave-packets, as in Eq. (\ref{psi}), and thin cyan and magenta curves are the corresponding flavor eigenstates; flavor oscillations occur where mass eigenstates overlap, see Appendix \ref{A} for technical details.

Because each interaction involves two flavor-mixed particles, the system is described by a two-particle wave-function, which has four components in the flavor and mass bases, namely
\beq
\ket{f\!f}\equiv \left(\begin{array}{c}
\alpha\alpha \\
\alpha\beta \\
\beta\alpha \\
\beta\beta
\end{array} \right)
\equiv
\left(\begin{array}{c}
\alpha_1\alpha_2({\bf x}_1,{\bf x}_2, t) \\
\alpha_1\beta_2({\bf x}_1,{\bf x}_2, t) \\
\beta_1\alpha_2({\bf x}_1,{\bf x}_2, t) \\
\beta_1\beta_2({\bf x}_1,{\bf x}_2, t)
\end{array} \right),
\label{ff}
\eeq
and
\beq
\ket{mm}\equiv \left(\begin{array}{c}
hh \\
hl \\
lh \\
ll
\end{array} \right)
\equiv
\left(\begin{array}{c}
h_1h_2({\bf x}_1,{\bf x}_2, t) \\
h_1l_2({\bf x}_1,{\bf x}_2, t) \\
l_1h_2({\bf x}_1,{\bf x}_2, t) \\
l_1l_2({\bf x}_1,{\bf x}_2, t)
\end{array} \right),
\label{mm}
\eeq
respectively, where the subscripts denote particle 1 and particle 2. Note that when the particles 1 and 2 are far apart (before or after an interaction), a two-particle wave-function is separable, being a direct product of one-particle ones: $\ket{f_if_j}=\ket{f_i({\bf x}_1,t)}\otimes \ket{f_j({\bf x}_2, t)}$, where $i=\alpha,\beta$, ~$j=\alpha,\beta$ and  $\ket{m_pm_q}=\ket{m_p({\bf x}_1,t)}\otimes \ket{m_q({\bf x}_2, t)}$, where $p=h,l$, ~$q=h,l$. The two-particle flavor and mass eigenstates are related as before, 
\beq
\ket{f\!f}=U_2\ket{mm}, 
\eeq
where the unitary matrix is
\rem{
\beq
U_2 \equiv U\otimes U = \left(
\begin{array}{cccc}
  \cos^2\theta  &  \cos\theta \ \sin\theta  &  \cos\theta \ \sin\theta  &  \sin^2\theta  
\\
 - \cos\theta \ \sin\theta  &  \cos^2\theta  & - \sin^2\theta  &  \cos\theta \ \sin\theta  
\\
 - \cos\theta \ \sin\theta  & - \sin^2\theta  &  \cos^2\theta  &  \cos\theta \ \sin\theta  
\\
  \sin^2 \theta  & - \cos\theta \ \sin\theta  &  - \cos\theta \ \sin\theta  &  \cos^2\theta  
\\
\end{array}
\right)
\eeq
}
\beq
U_2 \equiv U\otimes U = \left(
\begin{array}{cccc}
  c^2 & cs  &  cs  & s^2  
\\
 -cs & c^2 & -s^2 & cs
\\
 -cs & -s^2 & c^2 & cs 
\\
  s^2 & -cs &  -cs & c^2
\\
\end{array}
\right),
\eeq
in which $c=\cos\theta$ and $s=\sin\theta$. For simplicity, we will restrict further study to one-dimensional motion of particles. 

The evolution of the system at hand is described by the two-particle two-component Schr\"odinger equation. In the mass basis, it reads
\beq
i\hbar\partial_t\ket{mm(x_1,x_2,t)}=(H^{\rm free}+H^{\rm grav}+V)\ket{mm(x_1,x_2,t)}.
\label{SchE}
\eeq
Here the free particle Hamiltonian 
\beq
H^{\rm free}=\left(\begin{array}{cccc}
H^{\rm free}_{hh} & 0 & 0 & 0 \\
0 & H^{\rm free}_{hl} & 0 & 0 \\
0 & 0 & H^{\rm free}_{lh} & 0 \\
0 & 0 & 0 & H^{\rm free}_{ll} 
\end{array}\right)
\eeq
satisfies energy conservation, where
\begin{eqnarray*}
H^{\rm free}_{hh} & = & -\hbar^2\partial^2_{x_1x_1}/2m_h -\hbar^2\partial^2_{x_2x_2}/2m_h, \\
H^{\rm free}_{hl} & = &-\hbar^2\partial^2_{x_1x_1}/2m_h -\hbar^2\partial^2_{x_2x_2}/2m_l -\Delta mc^2, \\
H^{\rm free}_{lh} & = &-\hbar^2\partial^2_{x_1x_1}/2m_l -\hbar^2\partial^2_{x_2x_2}/2m_h -\Delta mc^2, \\
H^{\rm free}_{ll} & = &-\hbar^2\partial^2_{x_1x_1}/2m_l -\hbar^2\partial^2_{x_2x_2}/2m_l -2\Delta mc^2
\end{eqnarray*}
and $\Delta m\equiv m_h-m_l$. Gravity enters via 
\beq
H^{\rm grav}=\left(\begin{array}{cccc}
H^{\rm grav}_{hh} & 0 & 0 & 0 \\
0 & H^{\rm grav}_{hl} & 0 & 0 \\
0 & 0 & H^{\rm grav}_{lh} & 0 \\
0 & 0 & 0 & H^{\rm grav}_{ll} 
\end{array}\right),
\eeq
where
\begin{eqnarray*}
H^{\rm grav}_{hh} & = & m_h\phi(x_1)+m_h\phi(x_2), \\
H^{\rm grav}_{hl} & = & m_h\phi(x_1)+m_l\phi(x_2), \\
H^{\rm grav}_{lh} & = & m_l\phi(x_1)+m_h\phi(x_2), \\
H^{\rm grav}_{ll} & = & m_l\phi(x_1)+m_l\phi(x_2)
\end{eqnarray*}
and $\phi(x)$ is an arbitrary gravitational potential.

The interaction matrix is diagonal in the flavor basis,
\beq
\tilde V=
 \left(\begin{matrix}
V_{\alpha\alpha} & 0 & 0 & 0 \\
0 & V_{\alpha\beta} & 0 & 0 \\
0 & 0 & V_{\beta\alpha} & 0 \\
0 & 0 & 0 & V_{\beta\beta} \\
\end{matrix}\right),
\eeq
where $V_{\beta\alpha}=V_{\alpha\beta}$ for indistinguishable particles. In the mass basis, we have
\beq
V = U_2^\dagger \tilde V U_2 =\left(
\begin{array}{cccc}
A & E & E & D 
\\
E & C & D & F  
\\
E & D & C & F
\\
D & F & F & B
\\
\end{array}
\right) ,
\label{V}
\eeq
where the hermitian conjugate $U_2^\dagger=U_2^{-1}=U_2^T$ for the real-valued unitary matrix and
\begin{eqnarray*}
A & = & \textstyle\frac{1}{8} \left[ 3 V_{\alpha\alpha} + 2 V_{\alpha\beta} + 3 V_{\beta\beta} 
+ 4 (V_{\alpha\alpha} - V_{\beta\beta}) \cos2\theta 
\right.\\ & &{} \left. 
+ (V_{\alpha\alpha} - 2 V_{\alpha\beta} + V_{\beta\beta}) \cos4\theta \right] ,
\\
B & = & \textstyle\frac{1}{8} \left[ 3 V_{\alpha\alpha} + 2 V_{\alpha\beta} + 3 V_{\beta\beta}  
- 4 (V_{\alpha\alpha} - V_{\beta\beta} ) \cos2\theta 
\right.\\ & &{} \left. 
+ (V_{\alpha\alpha} - 2 V_{\alpha\beta} + V_{\beta\beta}) \cos4\theta \right] ,
\\
C & = & \textstyle\frac{1}{8} \left[ V_{\alpha\alpha} + 6 V_{\alpha\beta} + V_{\beta\beta} 
- (V_{\alpha\alpha} - 2 V_{\alpha\beta} + V_{\beta\beta}) \cos4\theta \right] ,
\\
D & = & \textstyle\frac{1}{4} \left[ V_{\alpha\alpha} - 2 V_{\alpha\beta} + V_{\beta\beta} \right] \sin^2 2\theta ,
\\ 
E & = & \textstyle\frac{1}{4} \left[ V_{\alpha\alpha} - V_{\beta\beta} 
+ (V_{\alpha\alpha} - 2 V_{\alpha\beta} + V_{\beta\beta}) \cos2\theta \right] \sin2\theta, 
\\
F & = & \textstyle\frac{1}{4} \left[ V_{\alpha\alpha} - V_{\beta\beta}
- (V_{\alpha\alpha} - 2 V_{\alpha\beta} + V_{\beta\beta}) \cos2\theta \right] \sin2\theta ,
\end{eqnarray*}
If the particles are distinguishable, one should make the substitution $2 V_{\alpha\beta}\to V_{\alpha\beta}+V_{\beta\alpha}$ in the above equations. Since trace is invariant under a unitary transformation, $\textrm{Tr}(V)=V_{\alpha\alpha} + 2 V_{\alpha\beta} + V_{\beta\beta}$; also useful is $\sum_{i,j}V^2_{ij}=V_{\alpha\alpha}^2 + 2 V_{\alpha\beta}^2 + V_{\beta\beta}^2$.

The physics represented by the $V$-matrix is easy to understand. There are four different interaction combinations (input channels): $hh\to\dots$, $hl\to\dots$, $lh\to\dots$, and $ll\to\dots$ in a statistical ensemble of indistinguishable particles interacting with each other, labeling of one particle to be the ``first'' one and the other to be the ``second'' is completely arbitrary, hence the states $hl$ and $lh$ describe the same statistical representation; nevertheless, we treat them separately for the sake of generality). In each of these interactions, there are four different outcomes (output channels): $\dots\to hh$, $\dots\to hl$, $\dots\to lh$, and $\dots\to ll$. Thus, the $V$-matrix `sandwiched' between initial and final states gives all 16 $S$-matrix elements, $S_{fi}=\, _f\!\bra{mm}V\ket{mm}_i$, where $i,f$ take the values $1,2,3,4\equiv (hh),(hl),(lh),(ll)$. More explicitly, for a given target particle $t$ and a projectile particle of species $s$, one has $S_{(s_it_i)(s_ft_f)}=\bra{m_{s_f}m_{t_f}} V\ket{m_{s_i}m_{t_i}}$ with $s_i$ and $s_f$ being the initial and final states of the projectile particle, $t_i$ and $t_f$ being those of the target particle. For example $S_{12}\equiv S_{(hh)(hl)}=\bra{hl}V\ket{hh}$ and corresponds to $hh\to hl$. 

There are two types of processes: (i) {\em elastic scatterings} in which the system composition does not change (e.g., $hh\to hh,\ hl\to lh$, etc.) and (ii) {\em mass eigenstate conversions} in which the composition changes (e.g., $hh\to hl,\ ll\to lh$, etc.). The diagonal elements $V_{ij}$ correspond to pure elastic scattering. Two off-diagonal elements, 23 and 32, describe `mass exchange', but they contribute to scattering as well if the particles are indistinguishable. All other elements represent conversion of one or two mass eigenstates. The total energy and momentum must be conserved in all processes. The energy-momentum conservation in elastic scattering is trivial, so we skip it. Conversions are different. Transitions in which a heavy eigenstate is converted into a light one go with the increase of kinetic energy and thus have no threshold. The opposite ones, where $l$ is converted into $h$, have a threshold $\Delta m c^2=(m_h-m_l)c^2$ and can only occur if kinematically allowed, i.e., if the initial kinetic energy of the interacting eigenstates is greater than the threshold. 

Interestingly, there is a set of parameters, for which the $S$-matrix elements for elastic interactions vanish identically but the conversion amplitudes (off-diagonal elements) do not. Indeed, (i) the diagonal matrix elements, Eq. (\ref{V}), namely $A,\ B,\ C$ contribute to the total elastic scattering cross-section, $\sigma_{\text{scat}}$; (ii) the off-diagonal `mass exchange' matrix elements $V_{23}=V_{32}=D$ also contribute to scattering in a statistical ensemble sense, if particles are indistinguishable; and (iii) the remaining elements $E,\ F$ and $V_{14}=V_{41}=D$ contribute to the total conversion cross-section, $\sigma_{\text{conv}}$. It is easy to see that one can have $\sigma_{\text{scal}}=0$ simultaneously with $\sigma_{\text{conv}}\not=0$. First, scatterings like $lh\to lh$ and $hl\to lh$ vanish if $C=D=0$, which requires that $V_{\alpha\beta}=V_{\beta\alpha}=0$, i.e., different flavors do not interact with each other, and also that $V_{\beta\beta}=-V_{\alpha\alpha}$. Second, scattering channels $hh\to hh$ and $ll\to ll$ vanish if $A=B=0$, which additionally requires maximal mixing, $\theta=\pi/4$. Thus, the matrix $V$ becomes 
\beq 
V= V_{\alpha\alpha}\left(\begin{matrix}
0 & 1 & 1 & 0 \\
1 & 0 & 0 & 1 \\
1 & 0 & 0 & 1 \\
0 & 1 & 1 & 0 \\
\end{matrix}\right)
\label{Vc}
\eeq
and $V_{\alpha\alpha}$ is the only independent matrix element. Thus, $V_{\text{scat}}=0$ (diagonal terms) and $V_{\text{exchange}}=0$ (i.e.,$V_{23}$ and $V_{32}$ terms, which play a role of scattering in a statistical ensemble) identically and $V_{\text{conv}}\not=0$, i.e., conversions can occur even if the gas of mixed particles has vanishing elastic scattering $S$-matrix elements. 

The $S$-matrix elements  $S_{(s_it_i)(s_ft_f)}$ are used to compute interaction cross-sections in the usual way \citep{LL}. Appendix {\ref{B} briefly discusses the scattering standard theory and presents some useful results. The scattered wave function can be expanded in angular momentum (or, equivalently, the impact parameter) as
\beq
\psi=\sum_{l=0}^\infty S_{(s_it_i)(s_ft_f)}^{(l)}P_l(\cos\theta)R_{l}(r),
\eeq
where $P_l$ are Legendre polynomials, $R_{l}(r)$ are radial functions being the solution of the radial part of the Schr\"odinger equation with a given scattering potential $V(r)$ and $S_{(s_it_i)(s_ft_f)}^{(l)}$ are partial $S$-matrix amplitudes of the processes $(s_it_i)\to (s_ft_f)$ for a given $l$. The elastic scattering [i.e, $(s_it_i)\to (s_it_i)$] cross-sections and the conversion [i.e., $(s_it_i)\to (s_ft_f)$, where $(s_it_i)\not= (s_ft_f)$] cross-sections are, see Eqs. (\ref{sigma-ii}), (\ref{sigma-ii}), 
\bea
\sigma_{(s_it_i)\to(s_it_i)} &=& \frac{\pi}{k_i^2}\sum_{l=0}^\infty(2l+1)\left|1-S_{(s_it_i)(s_it_i)}^{(l)} \right|^2, \\
\sigma_{(s_it_i)\to(s_ft_f)} &=& \frac{\pi}{k_i^2}\sum_{l=0}^\infty(2l+1)\left| S_{(s_it_i)(s_ft_f)}^{(l)} \right|^2.
\eea
where $k_i=p_i/\hbar$ is the initial wave-number in the center of mass frame.

In general, the cross-sections depend on the shape of the scattering potential, as well as particle momentum, $k=p/\hbar$ and angular momentum, $l$. However, for slow particles and sufficiently well-localized potentials $r_0 p/\hbar=r_0k\ll1$, $r_0$ being the characteristic size of the potential, the partial amplitudes with large angular momentum $l$ are small compared to the $l=0$ term. Thus, it is enough to keep the leading $l=0$ term in Eqs. (\ref{sigma-scat}), (\ref{sigma-conv}). Therefore
\bea
\sigma_{(s_it_i)\to(s_it_i)} &=& \frac{\pi \hbar^2}{p_{s_i}^2}\left|1-S_{(s_it_i)(s_it_i)} \right|^2,
\label{sigma-scat}\\
\sigma_{(s_it_i)\to(s_ft_f)} &=& \frac{\pi \hbar^2}{p_{s_i}^2}\left| S_{(s_it_i)(s_ft_f)} \right|^2.
\label{sigma-conv}
\eea
\rem{Note that this form of $\sigma$ manifestly satisfies the principle of detailed balance \citep{LL}, $\sigma_{i\to f}\,p_i^2=\sigma_{f\to i}\,p_f^2$, for the integral cross-sections of forward and reverse processes, and that $\sigma\propto1/v$.}

Two asymptotic cases worth noting. First, if the amplitude of conversions are much smaller than that of elastic scattering, then the cross-sections scale as follows, Eqs. (\ref{sigma-el-1}), (\ref{sigma-con-1}):
\bea
\sigma_{(s_it_i)\to(s_it_i)} &\propto& const.,\\
\sigma_{(s_it_i)\to(s_ft_f)} &\propto& 1/v_{s_i},
\eea
which are standard for $s$-wave scattering. Second, more interesting is the case of very efficient conversions (i.e., when the elastic $S$-matrix elements vanish). Then all the the cross-sections scale in the same way, Eqs. (\ref{sigma-el-2}), (\ref{sigma-con-2}):
\beq
\sigma_{(s_it_i)\to(s_it_i)} \sim \sigma_{(s_it_i)\to(s_ft_f)} \propto 1/v_{s_i}^{2}
\label{sigma-all}
\eeq
with an additional constraint that the elastic scattering cross-section is equal to the total cross-section of all conversions:
\beq
\sigma_{(s_it_i)\to(s_it_i)} = \sum\nolimits'_{(s_ft_f)} \sigma_{(s_it_i)\to(s_ft_f)}. 
\label{sigma-sum}
\eeq

For completeness, we also present a useful parameterization of the cross-section matrix, which can easily be implemented in numerical models:
\beq
\sigma_{(s_it_i)(s_ft_f)}=\sigma_0 \left(\frac{v_0}{v_{s_i}}\right)^\alpha B_{(s_it_i)(s_ft_f)},
\label{sigma-num}
\eeq
where $\sigma_0$ and $v_0$ are common normalization parameters and the elements of the $B$-matrix are proportional to the squares of the properly normalized matrix elements of the $V$-matrix, Eq. (\ref{V}) together with Eqs. (\ref{sigma-scat}) and (\ref{sigma-conv}), and simply enforce relative strengths of interaction channels that can occur. For instance, for the conversion-dominated interactions, the $B$-matrix follows from Eqs. (\ref{Vc}), (\ref{sigma-all}), (\ref{sigma-sum}) to be
\beq
B=\left(\begin{matrix}
2 & 1 & 1 & 0 \\
\Theta & 1+\Theta & 0 & 1 \\
\Theta & 0 & 1+\Theta & 1 \\
0 & \Theta & \Theta & 2\Theta \\
\end{matrix}\right),
\eeq
where $\Theta=\Theta(E_{s_ft_f})$ is the Heaviside function which ensures that the process is kinematically allowed (i.e., negative final kinetic energy, $E_{s_ft_f}<0$, means the process cannot occur), where $E_{s_ft_f}$ is equal to the initial kinetic energy of the particles in their center of mass frame plus $\Delta E$, which is $-\Delta mc^2$ for each $l\to h$ conversion, or $+\Delta mc^2$ for each $h\to l$ conversion; thus, for example, for $hl\to lh$, $\Delta E=0$  and for in the process $hh\to ll$, $\Delta E=2\Delta mc^2$.

\section{Kinematics of interactions}

Let us consider an illustrative example of interaction of $h$ and $l$ belonging to two different particles. As we mentioned before, we consider one-dimensional motion, for simplicity. Let us consider $hl\to ll$ conversion and we assume here that the inverse process $hl\to hh$ is kinematically forbidden. Before the interaction, the mass eigenstates propagate along geodesics which are different (because the eigenstates have different velocities) and localized in space (because the eigenstates are trapped). In the center of mass frame the momentum and energy conservations are $p_h+p_l=0=p_l'+p_l'$ and $(m_h^2c^4+p_h^2c^2)^{1/2}+(m_l^2c^4+p_l^2c^2)^{1/2}=(m_l^2c^4+{p'_l}^2c^2)^{1/2}+(m_l^2c^4+{p'_l}^2c^2)^{1/2}$, where `prime' means `after scattering', that is 
\beq
(m_h^2c^4+p_l^2c^2)^{1/2}+(m_l^2c^4+p_l^2c^2)^{1/2}=2(m_l^2c^4+{p'_l}^2c^2)^{1/2}.
\label{cons}
\eeq
and we remind here that the incoming particles are non-relativistic, $p_h, p_l\ll m_hc,m_lc$. If  $m_h\gg m_l$, the outgoing mass eigenstates are relativistic
\beq
p'_l\simeq\frac{c}{2}(m_h^2 +2m_h m_l -3m_l^2)^{1/2}\simeq \frac{m_hc}{2}.
\label{pl-prime}
\eeq
Alternatively, if the masses are degenerate $m_l\simeq m_h\simeq m$ and $\Delta m/m\ll 1$, then 
\bea
\Delta v  = v'-v &\simeq& \left[(\Delta m/m)c^2+v^2\right]^{1/2}-v
\nonumber\\
&\simeq&
\left\{\begin{array}{ll}
\frac{1}{\sqrt{2}}v_k, & \textrm{ if } v\ll v_k,\\
\frac{1}{4}v_k^2/v, & \textrm{ if } v\gg v_k,
\end{array}
\right.
\label{deltav-deg}
\eea
where we used that the velocities of $h$ and $l$ are also comparable, because $p_h= p_l$ in the center of mass frame; hence $v=p/m$ and $v'=p'/m$. Here we also introduced the ``kick velocity'', $v_k=c(2\Delta m/m)^{1/2}$ --- this is the velocity a heavy eigenstate at rest gets upon conversion into a light eigenstate, provided the recoil velocity of the scatterer is vanishing. Thus, after the interaction, the mass eigenstates propagate along new geodesics, and if $v_l'$ is greater than the escape velocity of the potential, $v_l'>v_{\rm esc}$, then both $\ket{m_l}$-eigenstates escape from the potential well. Alternatively, if elastic scattering occurs, $hl\to hl$ or $hl\to lh$, then the kinetic energy does not change and the eigenstates remain trapped. 

Therefore, upon any interaction involving the $hl\to ll$ process the amplitude of the heavy eigenstate decreased irreversibly and both eigenstates can become unbound. Though the total probability remains unity, the probability to detect the particle (an electron neutrino, for example) inside the potential has decreased and the probability of its detection somewhere outside has become larger. Of course, the overall energy is conserved: the light eigenstate climbs up the potential and loses energy (e.g., a massless particle is redshifted). By repeating this cycle, one can further decrease the amplitude of the trapped eigenstates. Colloquially speaking, the particles  ``evaporate'' from the potential well.

\section{Evolution of a two-particle system}

Here we show how two stable flavor-mixed particles, which are trapped in a gravitational potential and scatter off each other from time to time, gradually escape --- or ``evaporate'' --- from it. More precisely, the probability to detect the particles inside the potential decreases with time and the probability of their detection elsewhere increases. Such ``evaporation'' is a result of mass eigenstate conversions in which a heavier eigenstate converts into a lighter one, thus adding kinetic energy to the scattered particles. We emphasize that the phenomena of $m$-conversion and quantum evaporation are not related in any way to particle decays or other reactions, quantum tunneling and such. 

To illustrate the evaporation effect, we numerically solve the two-particle two-component Schr\"odinger equation, Eq. (\ref{SchE}). To ease numerical computations, we chose a model potential with strong screening, $\phi(x)=\phi_0e^{-(x/x_g)^{10}}(1-(x/x_g)^2)$, where $\phi_0<0$ (meaning that the potential is attractive) determines its depth and $x_g$ sets its size ($x_g\sim4$ in computational units). The interaction potential is given by Eq. (\ref{V}). Interactions of particles occur via a $\delta$-function potential, i.e., $V_{\alpha\alpha},V_{\alpha\beta},V_{\beta\beta}\propto \delta(x_1-x_2)$, which is numerically represented by $V_0\,e^{-[(x_1-x_2)/x_v]^2}(1+[(x_1-x_2)/x_v]^2)^{-1}$, where $V_0>0$ and $x_v\sim0.1$; the actual shape of $V(x_1-x_2)$ does not significantly affect the results so long as $x_v$ is small enough. The relative strengths are chosen to be $V_{\alpha\alpha}:V_{\alpha\beta}:V_{\beta\beta}=2:1:-2$, the mixing angle is $\theta=\pi/6$ and the masses are chosen to be degenerate, $\Delta m/m_h=0.15$. The initial wave-function components are taken to be gaussian wave-packets. 

Now we present exact numerical solutions of the Schr\"odinger equation for a pair of mixed particles. In order to simplify representation of the four-component two-dimensional time-dependent wave-function, Eq. (\ref{mm}), we compute probability densities of mass eigenstates for each particle (denoted by a subscript) as follows
\begin{subequations}
\bea
hh|_1(x,t)&\equiv&\int |hh(x,x_2,t)|^2\ dx_2,
\label{hh1}\\
hh|_2(x,t)&\equiv&\int |hh(x_1,x,t)|^2\ dx_1
\label{hh2}
\eea
\end{subequations}
and similarly for other components. Since the particles are indistinguishable, we define the total probability density of the heavy and the light mass eigenstates as
\begin{subequations}
\bea
h(x,t)&=&hh|_1+hh|_2+hl|_1+lh|_2,
\label{hxt}\\
l(x,t)&=&hl|_2+lh|_1+ll|_1+ll|_2.
\label{lxt}
\eea
\end{subequations}

\begin{figure}
\includegraphics[width=80mm]{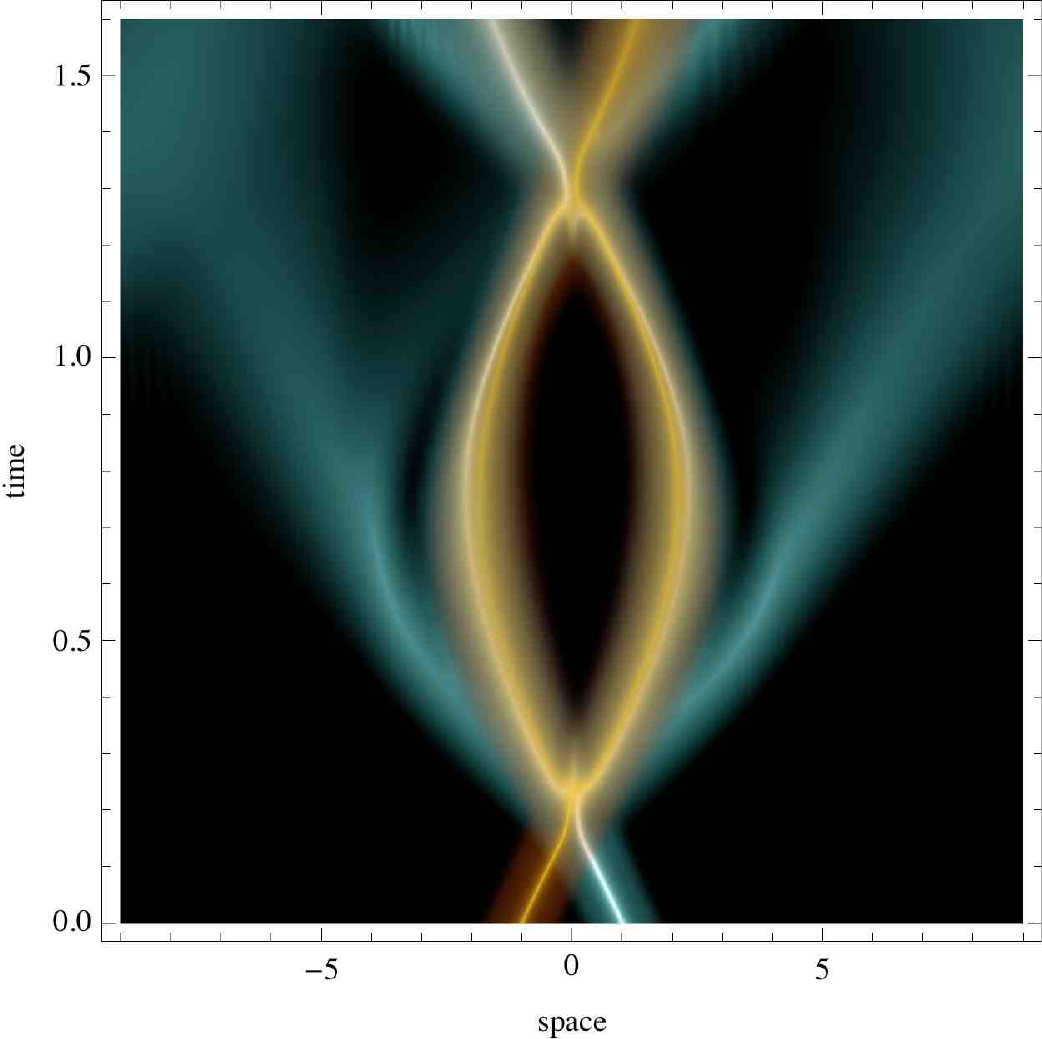}
\caption{Space-time diagram of two interacting mass eigenstates in a gravitational potential well showing the effect of $m$-conversions and particle evaporation, obtained by direct numerical solution of a two-particle two-component Schr\"odinger equation. The probability densities of light (cyan) and heavy (orange) mass eigenstates belonging to different particles are shown; yellow color originates from color blending of cyan and orange. The potential is localized between $x\sim-4$ and $+4$ (in computation units). At $t=0$ the system consists of $h_1$ and $l_2$ and both are trapped. During the evolution, each collision produces forward and reflected wave-packets of all possible mass eigenstates; those corresponding to conversions escape to infinity in the form of light mass eigenstates.
\label{Aevap}}
\end{figure}

\begin{figure*}
\includegraphics[width=170mm]{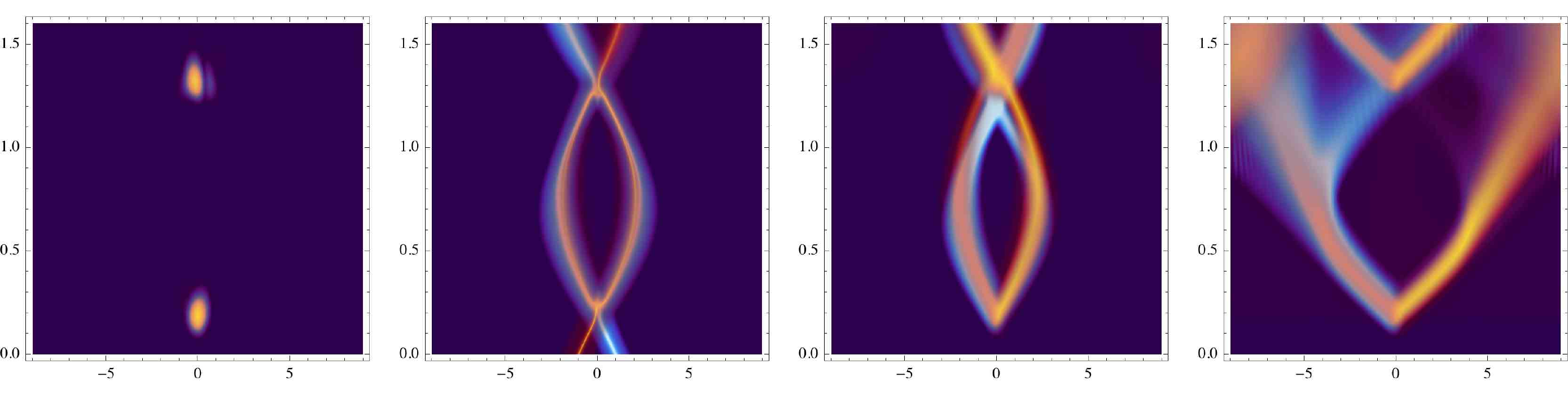}
\caption{Space-time diagrams, as in Fig. (\ref{Aevap}), but for the separate wave-function components: $hh$ (left panel), $hl$ (second panel), $lh$ (third panel) and $ll$ (right panel). Colors denote different particles: orange is particle 1 and blue is particle two. Initially, only $hl$ component of the 2-particle wave-function is non-zero, meaning the presence of $h$ mass eigenstate of particle 1 and $l$ mass eigenstate of particle 2. Other mass eigenstates are produced upon interactions, which are seen as vertexes. See text for more details and discussion.
\label{Acomponents}}
\end{figure*}

As a first case, we consider the interaction of a heavy and a light mass eigenstates belonging to two different particles, $h_1+l_2\to\dots$, which illustrates the effect of quantum evaporation. Fig. \ref{Aevap} shows the space-time diagram of the probability density of a heavy (orange) and a light (cyan) mass eigenstates given by Eqs. (\ref{hxt}), (\ref{lxt}); yellow color originates from color blending in regions where both mass eigenstates propagate along very similar paths. Initially, there is only the heavy mass eigenstate of particle 1 located at $x=-1$ (in computational units) and the light mass eigenstate of particle 2 located at $x=1$, and both are moving toward each other. Both eigenstates are initially gaussian wave-packets with momenta small enough to be trapped in the gravitational potential well. In each collision, forward and reflected wave-packets of all possible mass eigenstates are produced and light mass eigenstates participating in and/or resulting from conversions escape to infinity.

To further elucidate the dynamics of the interactions, we show in Fig. (\ref{Acomponents}) the wave-function components, Eqs. ({\ref{hh1}), (\ref{hh2}), namely, $hh|_j$ (first panel), $hl|_j$ (second panel), $lh|_j$ (third panel) and $ll|_j$ (last panel), where $j=1,2$. Here we use different color coding: orange represents particle 1 (i.e., $j=1$) and blue represents particle 2. As one can see, at $t=0$ only the state $hl$ (second panel) is non-vanishing; orange shows the wave-packet $hl|_1$ (i.e., the heavy eigenstate of particle 1 -- the only heavy eigenstate initially present in the system) and blue is the wave-packet $hl|_2$ (i.e., the light eigenstate of particle 2 -- the only light eigenstate initially present in the system). The first interaction occurs at $t\sim 0.2$ and the second at $t\sim 1.3$. Note that several processes occur at each interaction. First, no propagating $hh$ wave-packets form, as is seen from the first panel, because such $hl\to hh$ conversions are kinematically forbidden. Second, standard elastic collisions  $hl\to hl$ occur in which both forward- and back-scattered wave-packets are produced, as is seen in the second panel. Third, elastic ``exchange'' $hl\to lh$ also occurs, as is seen in the third panel (as we mentioned earlier, if the particles are indistinguishable, this process is equivalent to elastic scattering). Here the wave-function $lh$-component, which was initially absent, appears at $t\sim0.2$ as a vertex because both forward- and backward-scattered wave-packets appear. After that, the wave-packets of both particles propagate but remain trapped in the potential, so they meet each other again at $t\sim1.1$ (blue, particle 2) and $t\sim 1.4$ (orange, particle 1). Note that although the wave-packet paths intersect, no interactions occur: the wave-packets belonging to the same particle do not self-interact but can only interfere. Finally, the fourth panel shows that light eigenstates are produced in conversions $hl\to ll$ seen as vertexes at $t\sim 0.2$ and $t\sim 1.3$. The velocities of these wave-packets exceed the escape velocity (controlled by the potential depth) so they leave the gravitational potential.

\begin{figure}
\includegraphics[width=85mm]{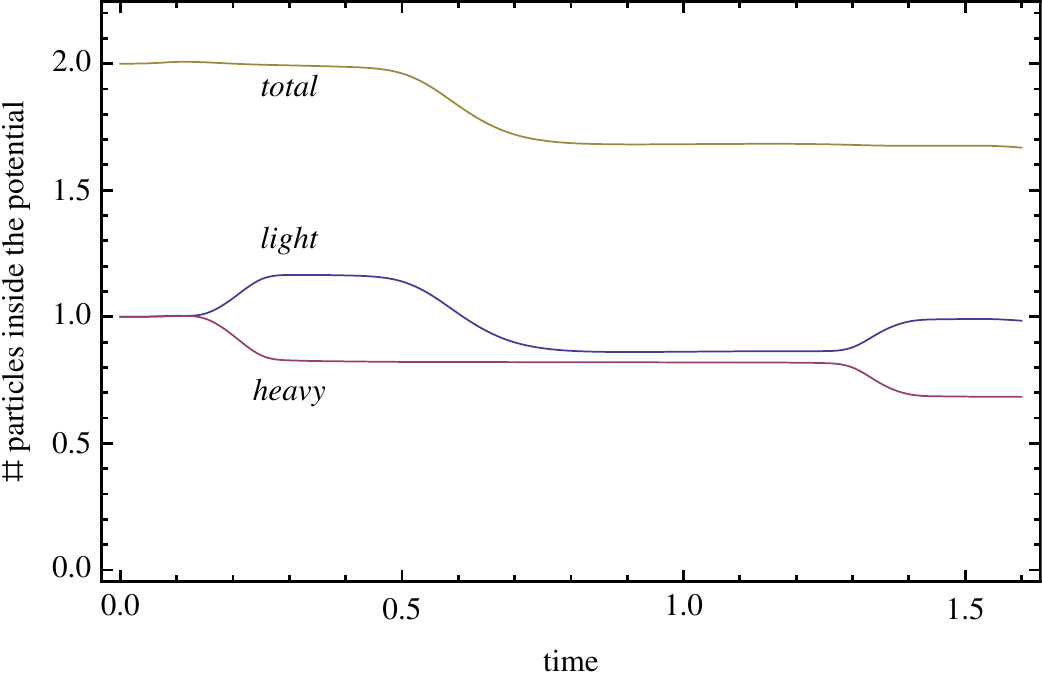}
\caption{Normalized expectation values of the number of particles confined inside the gravitational potential as a function of time, see Eqs. (\ref{Nh})--(\ref{N}). A heavy mass eigenstate is converted into a light mass eigenstate in each collision, at $t\sim0.2-0.3$ and $t\sim 1.3-1.4$. Soon after each collision/conversion, the light mass eigenstate escapes, thus decreasing the total confined mass.}
\label{Amass}
 \end{figure}

Finally, Fig. \ref{Amass} shows the expectation value of the number of particles inside the gravitational potential. To simplify comparison, we normalize them to the initial value as follows: 
\begin{subequations}
\bea
\hat n_h(t)&=&\frac{\int h(x,t)\,dx}{\int h(x,0)\,dx},
\label{Nh}\\
\hat n_l(t)&=&\frac{\int l(x,t)\,dx}{\int l(x,0)\,dx},
\label{Nl}\\
\hat n(t)&=&\hat n_h+\hat n_l,
\label{N}
\eea
\end{subequations}
where Eqs. (\ref{hxt}),  (\ref{lxt}) were used. One sees that a light mass eigenstate is produced in each collision ($t\sim0.2-0.3$ and $t\sim 1.3-1.4$) at the expense of the heavy eigenstate. Later, the light mass eigenstate escapes, thus decreasing the total mass inside.

\begin{figure}
\includegraphics[width=80mm]{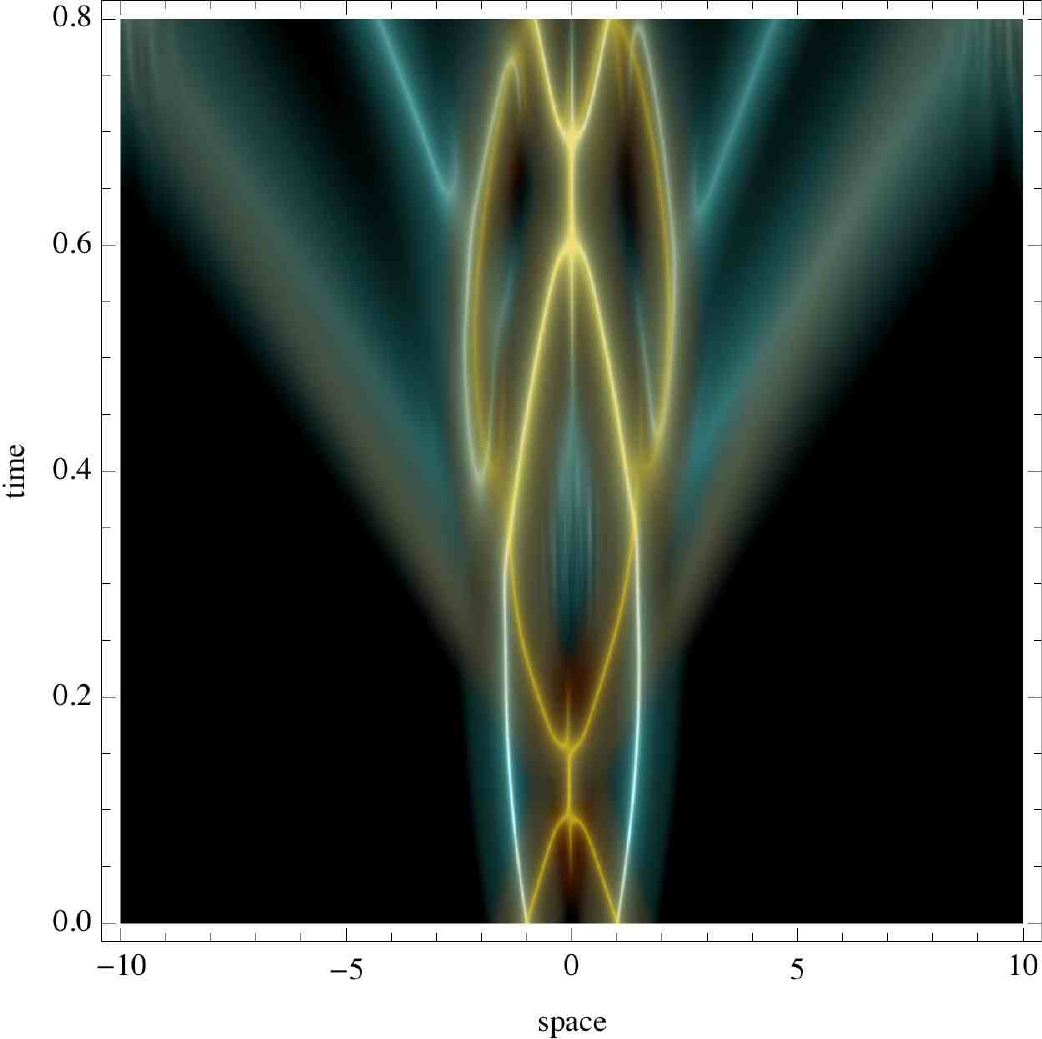}
\caption{Space-time diagram of two interacting flavor-mixed particles in a gravitational potential, similar to Fig. \ref{Aevap}. The probability densities of $l$ (cyan) and $h$ (yellow) mass eigenstates of both particles are shown. At $t=0$ the system consists of two particles of a particular flavor $\alpha$, each being the superposition of $h_1,\ l_1$ and $h_2,\ l_2$, and all the eigenstates are trapped inside a gravitational potential. At $t>0$, each collision produces forward and reflected wave-packets of all possible mass eigenstates and those with $v>v_\text{esc}$ escape to infinity.
\label{Bevap}}
\end{figure}

\begin{figure*}
\includegraphics[width=130mm]{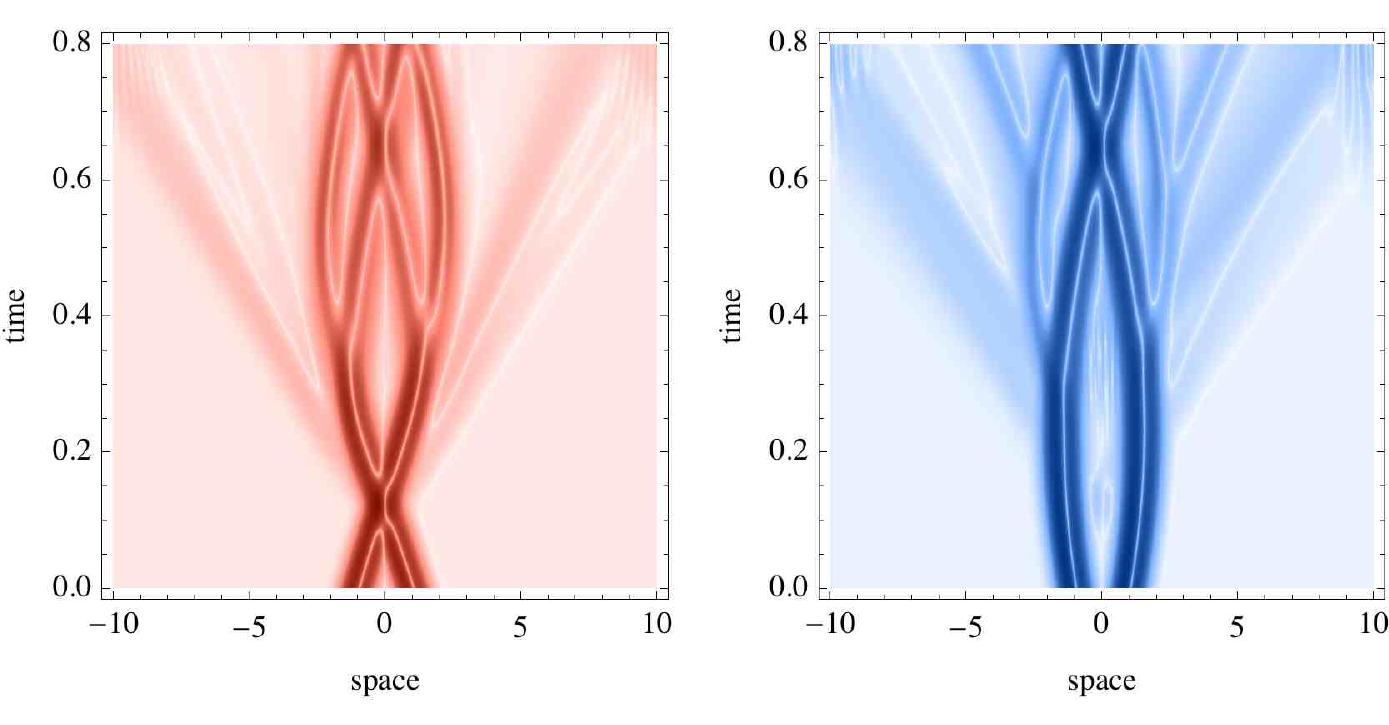} 
\caption{Evolution of mass eigenstates shown which are overlaid in Fig. \ref{Bevap}. Here, left panel (red) represents the $h(x,t)$ eigenstate given by Eq. (\ref{hxt}) and right panel (blue) is the $l(x,t)$, Eq. (\ref{lxt}).  
\label{Bstates}}
\end{figure*}

As a second case, we consider the full evolution of two flavor-mixed particles, each being a composition of both mass eigenstates. The essential difference of this case from the previous one is that all mass eigenstates of both particles are present. The initial state of the system is two flavor-mixed particles produced as flavor eigenstates at $x=-1$ and $x=1$ for particle 1 and 2, respectively. These particles are a coherent mixture of mass eigenstates propagating with different velocities: the heavy eigenstates move toward each other and the light ones move initially away from each other. This initial setup allows us  to separate the mass eigenstate interaction locations thus simplifying the analysis of the dynamics. Fig. \ref{Bevap} is analogous to Fig. \ref{Aevap} and shows the conversion of heavy mass eigenstates into light ones and their escape from the gravitational potential. Cyan and yellow colors here denote $l$ and $h$ mass eigenstates. To elucidate the dynamics, we also separate the mass eigenstates into different panels in Fig. \ref{Bstates}, which is otherwise identical to Fig. \ref{Bevap}.

\begin{figure*}
\includegraphics[width=170mm]{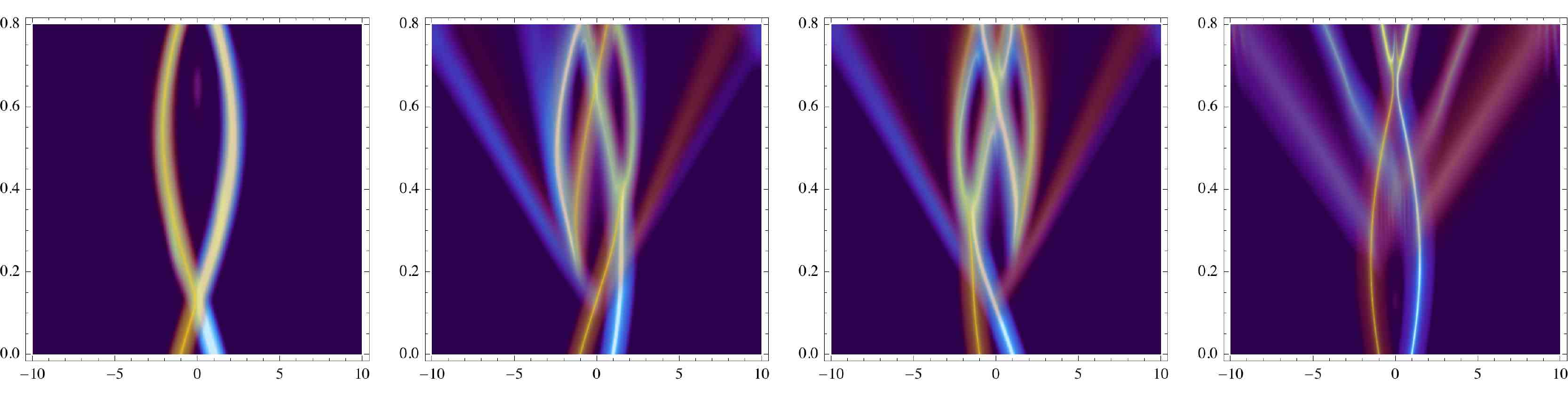}
\caption{Space-time diagrams, as in Fig. (\ref{Acomponents}) but for the two-particle case. Shown are wave-function components: $hh$ (left panel), $hl$ (second panel), $lh$ (third panel) and $ll$ (right panel); colors denote particles 1 (orange) and particle 2 (blue). Mass eigenstates conversions are seen as vertexes in these graphs (see text for detail).
\label{Bcomponents}}
\end{figure*}

\begin{figure}
\includegraphics[width=85mm]{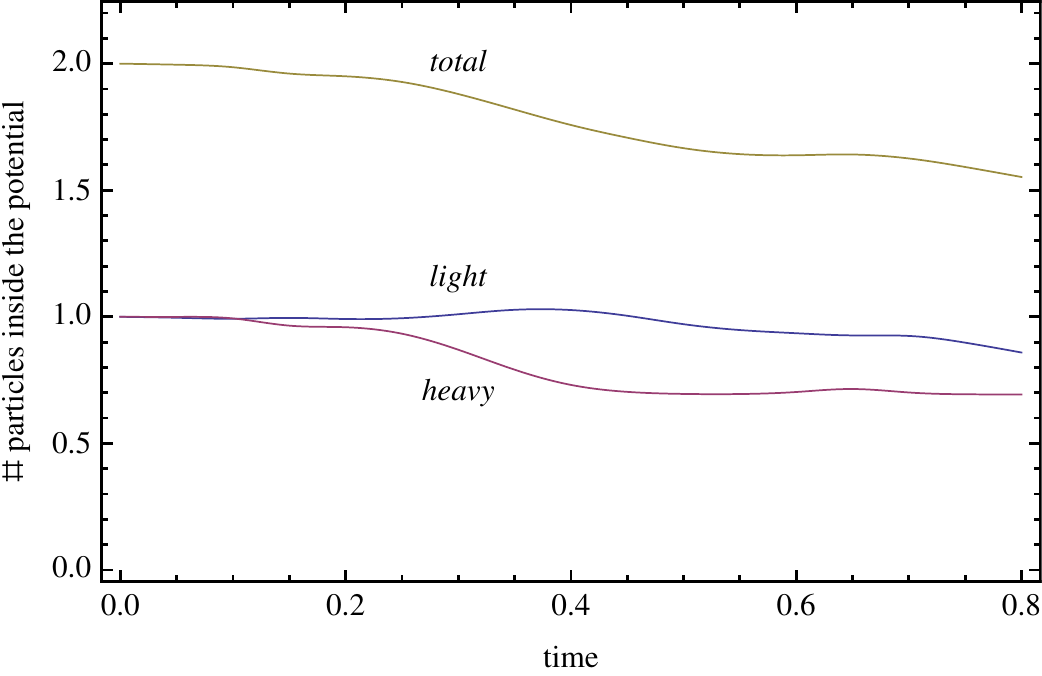}
\caption{Normalized expectation values of the number of particles confined inside the gravitational potential as a function of time, similar to Fig. \ref{Amass}. Any time a heavy mass eigenstate is converted into a light mass eigenstate in a collision, soon after that the light mass eigenstate escapes to infinity.}
\label{Bmass}
 \end{figure}

Figures \ref{Bcomponents} and \ref{Bmass} are similar to \ref{Acomponents} and \ref{Amass}. They show the evolution of the wave-function components and the number of particles inside the potential. From Fig. \ref{Bcomponents} one can see that $h$ mass eigenstates interact first (at $t\sim0.1$, first panel) to produce elastically scattered trapped $h$-states (panel one) and the outgoing $h$ and $l$-states via $hh\to hl$ and $hh\to lh$ (panels two and three, respectively), both have large enough velocities to escape ($h$ gets large $v$ by recoil). These escaping $h$-states interact on their way out (at $t\sim 0.4$) with the scattered trapped $h$-states to further produce escaping $h$- and $l$-states via the processes of conversion $hl\to ll$, $lh\to ll$ and ``exchange" $hl\to lh$, $lh\to hl$; trapped $h$-states are also produced at this time via inverse processes $hl\to hh$, $lh\to hh$. Such processes repeat later as well, e.g., at $t\sim0.6$. The amplitude of the direct $hh\to ll$ conversions is rather small for the chosen mixing angle and the values of $\tilde V$-matrix, so they are not visible in this figure. However, they are seen in Fig. \ref{Bmass} as the decrease of the mass of the heavy eigenstate at $t\sim 0.1$, when only $hh$ collision had occurred. The recoil velocity is larger in this process, hence the light eigenstate escape is fast. Overall, one can see from Fig. \ref{Bmass} that the particle evaporation is rapid and efficient in this case. 

We also note that the above examples are one-dimensional for illustration purposes. Whereas they captures all essential physics of the mixed-particle interactions, they cannot be used to evaluate interaction cross-sections for real three-dimensional world. The three-dimensional cross-sections are generally much smaller than the one-dimensional ones because the colliding particles have a huge phase space to miss each other.

\section{Asymptotic state, $t\to\infty$}

We demonstrated that evaporation of both light and heavy eigenstate can occur, which opens up a possibility of complete evaporation of both particles, which were initially trapped. What conditions are needed for this to occur? Here we present some general estimates; a dedicated analysis may be needed for a specific system. Let the initial composition of the trapped particle population be $n_{h,0}$ and $n_{l,0}$. For a single two-component particle of flavor $\alpha$, these are $n_{h,0}=\cos^2\theta$ and $n_{l,0}=\sin^2\theta$, and for a particle of flavor $\beta$, they are $n_{h,0}=\sin^2\theta$ and $n_{l,0}=\cos^2\theta$, as follows from Eq. (\ref{mix}). Note that in both cases $n_{h,0}+n_{l,0}=1$, i.e., there is exactly one particle in the system. If we consider a system of many particles, $n_{h,0}$ and $n_{l,0}$ must be multiplied by the number of particles. 

Let us also assume that the system is ``optically thin'', i.e., probability of particle interaction during one bounce is very small, so if a conversion occurred, the escaping eigenstate experiences no further interactions and just leaves the system for good. We also assume that only forward conversions ($h\to l$) can occur; inverse processes ($l\to h$) are kinematically forbidden. We consider indistinguishable particles and also assume that $v_k>v_\text{esc}$. These assumptions are very natural for non-relativistic mixed particles such as neutrinos (e.g., relic neutrinos from big bang) and some dark matter candidates because of their very small interaction cross-sections. 

The composition at $t>0$ is described by $n_h(t)$ and $n_l(t)$, which are governed by equations
\begin{subequations}
\bea
\dot n_h&=&-(\sigma_{hh} v)\,n_h^2-(\sigma_{hl}v)\,n_h n_l, 
\label{dot-nh}\\
\dot n_l&=&-(\sigma_{hl}v)\,n_h n_l,
\label{dot-nl}
\eea
\end{subequations}
where we also assumed, for simplicity, that the particle density is uniform throughout the system. Here $v$ is the relative velocity of two interacting eigenstates which are comparable for heavy and light eigenstates if $m_h\simeq m_l$. Here also $\sigma_{hh}$ is the total cross-section of the processes $hh\to hl, lh, ll$ and $\sigma_{hl}$ is the total cross-section of the processes $hl,lh\to ll$, hence $\sigma_{hh}\propto 2E^2+D^2$ and $\sigma_{hl}\propto 2F^2$, see Eqs. (\ref{V}), (\ref{sigma-conv}). Whereas the general solution to these equations has no simple analytical solution, the asymptotic state can be found as follows. From  Eqs. (\ref{dot-nh}), (\ref{dot-nl}):
\beq
\frac{d\, n_h}{d\, n_l}=\frac{\sigma_{hh}\,n_h}{\sigma_{hl}\,n_l}+1.
\eeq
This equation has a solution:
\begin{subequations}
\bea
\frac{n_h(t)}{n_{h,0}} &=& \left(\frac{n_{l,0}/n_{h,0}}{1-R}\right)\left(\frac{n_l(t)}{n_{l,0}}\right)
\nonumber\\
& &{} +\left(1-\frac{n_{l,0}/n_{h,0}}{1-R}\right)\left(\frac{n_l(t)}{n_{l,0}}\right)^R,
\eea
where $R=\sigma_{hh}/\sigma_{hl}\not=1$, and
\beq
\frac{n_h(t)}{n_{h,0}} = \frac{n_l(t)}{n_{l,0}}\left[1+\frac{n_{l,0}}{n_{h,0}}\,\ln\!\left(\frac{n_l(t)}{n_{l,0}}\right)\right], 
\eeq
\end{subequations}
if $R=1$. We still do not know $n_h(t)$ and $n_l(t)$, but we note that $h\to l$ conversions will occur as long as $n_h(t)\not=0$. Therefore, asymptotically, when $n_h(\infty)\to 0$, $n_l(\infty)\to n_{l,\infty}$ -- some constant value:
\begin{subequations}
\beq
\frac{n_{l,\infty}}{n_{l,0}}=\left[1-\frac{n_{h,0}}{n_{l,0}}(1-R)\right]^{\frac{1}{1-R}},
\eeq
which is valid for both $0\le R<1$ and $R>1$, and 
\beq
\frac{n_{l,\infty}}{n_{l,0}}=\exp\!\left(-\frac{n_{h,0}}{n_{l,0}}\right),
\eeq
\end{subequations}
if $R=1$. 

We now conclude that when the initial composition satisfies the inequality
\beq
\frac{n_{l,0}}{n_{h,0}}\le 1-\frac{\sigma_{hh}}{\sigma_{hl}},
\label{evap-condition}
\eeq
complete evaporation of mixed particles occurs, that is no particles will be left inside the gravitational well, $n_{h,\infty}=n_{l,\infty}=0$. Of course, the particles will be outside and traveling to infinity as light mass eigenstates only. This means that the flavor composition will be $n_\alpha : n_\beta=\sin^2\theta :\cos^2\theta$.

\section{Conversions in Minkovsky space}

It is also important to investigate interactions of the particles in free space when gravity is negligible. This regime is relevant, for example, for the flavor-mixed dark matter in the early universe before structure formation starts, and for the relic cosmological neutrinos when they eventually become non-relativistic but still too hot to be confined by the gravitational attraction of the the large scale structure.

As before, mass eigenstates of a mixed particle move as if they are normal particles with certain (unequal) velocities and masses. The key difference between free and gravitationally confined particles is how their wave-packets spread with time. Depending on the shape of the potential, the wave-packet of a trapped particle, generally, spreads slower than in free space or even contracts (e.g., near the turning points). In this case, the separation of mass eigenstates occurs rapidly and can be nearly perfect as $t\to\infty$, so one can treat these eigenstates independently. In contrast, the wave-packets widths of free particles grow linearly with time and so does the separation between them. Therefore, the wave-packets of the two mass eigenstates can remain partially overlapped as $t\to\infty$, and the effect may be very significant depending on particle masses. Particle interactions in this case will involve both mass eigenstates leading to suppression of mass-conversion amplitudes. For example, when mass eigenstate wave-packets perfectly overlap, each particle is in a specific flavor eigenstate, and interactions do not change particle flavors (and hence mass eigenstate composition) by definition of an eigenstate. 

Let us consider a non-relativistic mixed particle created at some moment of time $t=0$ at a position $x_0$ in a certain flavor eigenstate. It is a coherent superposition of mass eigenstates and each is described by a wave-packet, which we assume here to be gaussian:
\beq
\psi_j(x)=\frac{1}{(2\pi\Delta_0^2)^{1/4}}\exp\left[-\frac{(x-x_0)^2}{4\Delta_0^2}+i\frac{m_jv_jx}{\hbar}\right],
\label{psi0}
\eeq
where $\Delta_0$, $m_j$ and $v_j$ are the wave-packet width, mass and velocity and $j=h,l$. The first term describes a gaussian shape and the second term is simply the phase $ikx=i(p/\hbar)x=imvx/\hbar$. Note that $\Delta_0$ is the same for all mass eigenstates because the wave-packets must overlap completely at $t=0$ --- the particle is created in a well-defined flavor eigenstate everywhere (i.e., at any $x$). Here we consider a one-dimensional case. At any time $t>0$ the wave-packet $\psi_j(x,t)$ is given by the solution of the Schr\"odinger equation \citep{WPspread} for an initial state $\psi_j(x)\equiv\psi_j(x,0)$, i.e.,
\begin{widetext}
\bea
\psi_j(x,t)&=&\left(\frac{m_j}{2\pi i\hbar t}\right)^{1/2}\!\!\int_{-\infty}^{\infty}dx'\,
\exp\left[-\frac{m_j(x-x')^2}{2i\hbar t}\right]\psi_j(x') 
\nonumber\\
&=& \left[2\pi\left(\Delta_0+\frac{i\hbar t}{2m_j\Delta_0}\right)^2\right]^{-1/4}
\exp\left[-\frac{(x-x_0-v_jt)^2}{4\Delta_0^2+2i\hbar t/m_j}+\frac{i}{\hbar}\left(m_jv_jx-\frac{m_jv_j^2}{2}t\right)\right].
\label{psi}
\eea
\end{widetext}
The generalization of this result to three dimensions is straightforward: wave packet spreading occurs independently in each orthogonal Cartesian direction $x_j=(x,y,z)$. This can be seen from that the gaussian wave-packet in three dimensions is separable into a product of three one-dimensional gaussians, the Hamiltonian of a non-relativistic free particle is quadratic in momentum $p^2=p_x^2+p_y^2+p_z^2$, and the orthogonal components of ${\bf p}$ and ${\bf x}$ commute, $[p_i,x_j]=0$ if $i\not=j$. Thus the triple integral in $d{\bf x}'$ breaks down into three single integrals. The result is: the coordinates and velocities in Eq. (\ref{psi}) become vector quantities.

This wave-function describes motion of $j$-th eigenstate with velocity $v_j$ and the wave-packet spreading due to the momentum uncertainty, $\Delta p \Delta_0\simeq\hbar$. In general, the velocities $v_h,v_l$ are different so the wave-packets of different mass eigenstates tend to separate in time: the gaussian centroids separate as $\delta x(t)\sim(v_l-v_h)t\propto t$. On the other hand, the widths of the wave-packets also grow in time as $\Delta(t)\sim(\hbar/m_j\Delta_0)t\propto t$ as $t\to\infty$. Since both grow linearly in time at late times, there will always be a non-zero overlap of the mass eigenstates.

Interactions of mass eigenstates occur as follows. First, if the mass eigenstate wave-packets overlap completely, they both interact simultaneously as a flavor wave-function. This results in elastic scatterings only (flavor is conserved in interactions), because the interaction hamiltonian, $\tilde V$, is diagonal is flavor basis, and no $m$-process can occur. Second, in the opposite case of completely separated mass eigenstates, as in the case of trapping in a gravitational field discussed earlier, the interaction matrix is non-diagonal, so both elastic scattering and conversions do occur. Finally, if the mass eigenstates partially overlap, there are non-zero chances for the particle to interact along both scenarios. In particular, interactions as flavor eigenstates (i.e., non-separated mass eigenstates) is proportional to the overlap integral of the mass wave-packets. We calculate the overlap integral now. 

A wave-packet given by Eq. (\ref{psi}) can be written as $\psi_j(x,t)=A_j(x,t)e^{i\phi_j(x,t)}$, where $\phi_j$ is a real-valued phase and $A_j$ is the real-valued amplitude which determines the shape of the wave-packet. Since $A^2_j=\psi_j^*\psi_j$ we readily obtain:
\beq
A_j=\left[2\pi\Delta_j^2(t)\right]^{-1/4}\exp\left[-\frac{(x-x_0-v_jt)^2}{4\Delta_j^2(t)}\right],
\label{A}
\eeq
where the wave-packet width is
\beq
\Delta_j^2(t)=\Delta_0^2+\left(\frac{\hbar}{2m_j\Delta_0}\right)^2t^2.
\label{sig0}
\eeq
Note that $x_0$ and $\Delta_0$ are the same for both mass eigenstates because the particles are produced as flavor eigenstates, hence the mass eigenstate wave-packets completely overlap at $t=0$. If the particles form an ensemble in thermal equilibrium with some temperature $T$ --- the case that can be relevant to the early universe conditions --- the expression for $\Delta(t)$ can readily be generalized \citep{WPspread} to yield
\beq
\Delta_j^2(t)=\Delta_0^2+\left(\frac{\hbar^2}{4m_j^2\Delta_0^2}+\frac{kT}{m_j}\right)t^2.
\label{sig1}
\eeq

The overlap integral of two mass eigenstates, $h$ and $l$ is
\beq
I(t)=\int_{-\infty}^\infty \, A_h(x,t)A_l(x,t)\,dx,
\eeq
where $A_h$ and $A_l$ are given by Eq. (\ref{A}). This integral is easily calculated analytically to yield:
\beq
I(t)=\left(\frac{2 \Delta_h\Delta_l}{\Delta_h^2+\Delta_l^2}\right)^{1/2}\exp\left[-\frac{t^2(v_h-v_l)^2}{4\left(\Delta_h^2+\Delta_l^2\right)}\right],
\eeq
where $\Delta_h^2$ and $\Delta_l^2$ are given by Eqs. (\ref{sig0}) or (\ref{sig1}). It's easy to check that $I(0)=1$, that is the wave-packets overlap completely at $t=0$, and $1>I(t)>0$ at $t>0$. 

To estimate the rate of $m$-conversions, we look for the minimum overlap, i.e., for the asymptotic value of $I(t)$ as $t\to\infty$. The mass eigenstates need not have same momenta or energy, see discussion in \citep{AS09,Giunti01}, not those assumptions are Lorentz invariant. However, for the sake of simplicity, here we choose them to have the same momenta\footnote{Note that if the momenta of the mass eigenstates are different, then their wave-packets, Eq. (\ref{psi}), carry extra $x$-dependent phase even at $t=0$, hence the particle exhibits flavor oscillations through space.} $p$, so that $v_h=p/m_h$ and $v_l=p/m_l$. 

First, we consider the case with strong mass-degeneracy: $m_h\approx m_l\approx m$, $\Delta m\equiv m_h-m_l\ll m$. We obtain
\beq
I(\infty)\simeq1-\left(\frac{\Delta m}{m}\right)^2\xi+{\cal O}\left(\frac{\Delta m^3}{m^3}\right),
\eeq
where $\xi$ is a numerical factor of order unity. Indeed, if $T=0$, Eq. (\ref{sig0}) holds, hence
\beq
\xi=\frac{1}{4}+\frac{p^2\Delta_0^2}{2\hbar^2}\sim \frac{1}{4}+\frac{p^2}{2(\Delta p)^2}\sim {\cal O}(1),
\eeq
where $\Delta p$ is the momentum uncertainty and we used that $\Delta p\Delta x\simeq \hbar$ with $\Delta x\sim\Delta_0$ and that $\Delta p\sim p$ in collisions. In the opposite case when $T$ is large enough for the first term in the brackets in Eq. (\ref{sig1}) to be neglected, one has
\beq
\xi=\frac{1}{16}+\frac{p^2}{8mkT}\sim \frac{1}{16}+\frac{E_{\rm th}}{4kT}\sim {\cal O}(1),
\eeq
where $E_{\rm th}=p^2/2m\sim \frac{3}{2}kT/m$ is the thermal energy of a particle. Overall, one can see that the value of $I(\infty)$ is fairly insensitive to the model assumptions and the estimate 
\beq
I(\infty)\sim1-(\Delta m/m)^2
\eeq 
is robust.

Second, if the masses are non-degenerate, $m_h\gg m_l$, then 
\beq
I(\infty) \simeq \eta\left(\frac{m_l}{m_h}\right)^{1/2}+{\cal O}\left(\frac{m_l^{3/2}}{m_h^{3/2}}\right),
\eeq
where
\bea
\eta &=& \sqrt{2}\,\exp\!\left[-\frac{\Delta_0^2 p^2}{\hbar^2}\right]\left(1+\frac{4\Delta_0^2 kT m_h}{\hbar^2}\right)^{1/4}
\nonumber\\
&\sim&\sqrt{2}\,e^{-1}\left(1+4/3\right)^{1/4}\sim {\cal O}(1).
\eea
Thus, in this case the overlap is negligible,
\beq
I(\infty)\sim(m_l/m_h)^{1/2}\ll1.
\eeq

We have found that mass eigenstates can rapidly become well-separated in a gravitational field, where they propagate along significantly different geodesics, or in flat space-time, where the local gravitational fields are extremely weak, provided there masses are very different. However, if the mass eigenstates have degenerate masses and are propagating in Minkovsky space, their wave-packets spread much more rapidly than their centroids move apart. These mass eigenstates thus remain nearly perfectly overlapped at all times, $I(\infty)\simeq1$. Should it be identically unity, no conversions would occur. Due to the slight non-overlap, the conversion amplitude is small but nonzero, being a factor of $(\Delta m/m)^2$ smaller than the conversion amplitude in the case of complete separation of the wave-packets. Thus the conversion cross-section in flat space-time, being proportional to the amplitude squared, is much smaller than that when mass eigenstates are well-separated, e.g., in the presence of sufficiently strong gravitational field, thus
\beq
\sigma_{\rm conv}^{\rm fst}\sim (\Delta m/m)^4\sigma_{\rm conv},
\label{sigma-fs}
\eeq
if $\Delta m\ll m$ and $\sigma_{\rm conv}^{\rm fst}\sim\sigma_{\rm conv}$ otherwise.

\section{Implications}

There are interesting cosmological implications of the obtained results.

The first implication concerns with cosmological neutrinos. Neutrinos from the cosmic neutrino background (CNB) have recently become non-relativistic; their thermal velocities are $v_\text{th}\simeq 81(1+z)(\text{eV}/m_\nu)~\text{km s}^{-1}$ \citep{Wong11}, which is of the order of a few hundred to a thousand km/s, hence they can be trapped in dark matter halos of large galaxies and galaxy clusters \citep{LvZ13}. Scattering of neutrinos off matter, though weak (but it can be greatly enhanced by coherent effects \citep{Sh+82}), will result in their mass eigenstate conversions and escape. 

Detectors on Earth, if they will ultimately be able to detect CNB neutrinos, should see the fractional deviation from the uniform composition of order unity for upward vs. downward going relic neutrinos. Indeed, the non-relativistic neutrino-nucleon cross-section is $\sigma_0\simeq G_F^2E_\nu^2\simeq5\times10^{-56}(E_\nu/\text{eV})^2\text{ cm}^2$ with $G_F$ being the Fermi constant of weak interactions. Thus, for the heaviest species, assuming $E_\nu^2\simeq\Delta m_{23}^2\simeq 0.0027\text{ eV}^2$, we have $\sigma_0\simeq1.4\times10^{-58}\text{ cm}^2$. The effect of coherent scattering increases the cross-section tremendously \citep{Sh+82}: $\sigma_\nu\simeq\sigma_0 Z^2 N^2$, where $Z$ is the charge of atomic nuclei, $N\simeq n V_\lambda$ is the number of nuclei in the volume $V_\lambda\simeq(4\pi/3)\lambda_\text{dB}^3$, $n$ is the number density of nuclei and $\lambda_\text{dB}=h/(m_\nu v_\text{th})\sim 0.5\text{ cm}$ is the neutrino de Broglie wavelength at $z=0$ (note, it is independent of $m_\nu$ for CNB neutrinos). For Earth, $Z\simeq 25$, $n\simeq10^{23}\text{ cm}^{-3}$, so the CNB neutrino cross-section in Earth is $\sigma_{CNB}\simeq2\times10^{-10}\text{ cm}^2$. The characteristic number density of the coherent scatterers in Earth is $n_\lambda\simeq1/V_\lambda$ and the typical distance neutrinos travel in Earth is its diameter, $d\simeq 10^9\text{ cm}$, hence the `optical depth' of Earth for the CNB neutrinos is $\tau\simeq\sigma_{CNB}n_\lambda d\simeq0.4$, so the modifications to the relic neutrino composition and spectrum will be large. Interestingly, only objects like rocky planets are important for CNB conversions. For example, from the the same calculation for the Sun yields the negligible `conversion optical depth' $\tau\sim10^{-14}$. We emphasize that CNB distortions are strongest on Earth (for upward-moving neutrinos) and this is where we should look for them, but the effect of conversions on the present-day neutrino cosmology is vanishing, as we'll see in the next paragraph. It is also important to reiterate that $m$-conversions change their kinetic energy. Hence the energy distribution of neutrinos should have, upon conversions, three spectral peaks at energies around $\sqrt{\Delta m_{12}^2}$, $\sqrt{\Delta m_{23}^2}$ and $\sqrt{\Delta m_{13}^2}$ corresponding to conversions $m_2\to m_1$, $m_3\to m_2$ and $m_3\to m_1$, respectively. 

We can speculate about the ultimate fate of the CNB in the universe. As the age of the universe (i.e., the Hubble time) $t_H\to\infty$, only the lightest mass eigenstates, $m_1$, of neutrinos will be present. From the Pontecorvo-Maki-Nakagawa-Sakata matrix, using presently measured mixing angles, we predict the asymptotic flavor composition, $\nu_e : \nu_\mu : \nu_\tau$, to be $1 : (0.3+0.1\cos{\delta}) : (0.2-0.1\cos{\delta})$, where $\delta$ is the yet-unknown CP-violating phase. (These are simply the ratios of probabilities of different flavors for the given lightest mass-eigenstate $m_1$.) Because of the very small neutrino cross-section and the very low average density in the universe, it will take much longer than the current age of the universe, which is $\sim4\times10^{17}$~s, to achieve this asymptotic distribution. We can estimate this time for a galaxy like Milky Way (of mass $M\sim10^{12}M_\odot$, size $R\sim15$~kpc and containing about $10^{11}$ Sun-like stars), assuming that conversions are efficient in rocky planets only and that there is one such planet per star in the Galaxy, to be $t_\nu\sim1/(n_p\sigma_p v_p)$, where $n_p\sim3\times10^{-58}~\text{cm}^{-3}$ is the number density of planets in the Galaxy, $\sigma_p\sim10^{18}~\text{cm}^2$ is a typical Earth-like planet cross-section and $v_p\sim(GM/R)^{1/2}\sim2\times10^7~\text{cm~s}^{-1}$ is the characteristic velocity in the Galaxy. One obtains $t_\nu\sim5\times10^{32}~\text{s}$, which is about $10^{15}$ times the current age of the universe. 

In order for such a process to occur in the first place, the decoherence (i.e., the mass eigenstate separation) should be much faster than the time between successive scattering. The decoherence time for a gravitationally bound particles is the time-scale on which their geodesics diverge substantially. This is roughly the particle travel time through the gravitational potential. Upon conversion, the speed of a lighter neutrino eigenstate is roughly the speed of light, unless, they are highly mass-degenerate. But even then, the relative velocities of the secondary mass eigenstates should be of the order of the escape velocity from the halo, otherwise the secondaries will remain trapped. For a Milky Way-type halo, the escape velocity is a few hundred kilometers per second and its size is about a hundred kiloparsecs. Thus, the decoherence time is $t_d\sim 10^{16}~\text{s}$ or shorter. This timescale is much shorter than $t_\nu$, hence the $m$-conversions will take place. Note that this estimate of the decoherence time is good for trapped particles in general, not just neutrinos, so it is applicable to dark matter as well, which is discussed below.

The second implication is more speculative and deals with the recently proposed two-component dark matter \citep{M10,M13}. A number of dark matter candidates are flavor-mixed particles \citep{DMcandidates-rev}. If two or more mass eigenstate are stable and haven't decay into the lightest one, also and these particles can self-interact, then the conversions discussed in this paper will affect the composition, structure and dynamics of dark matter halos. The cold dark matter (CDM) paradigm correctly describes the large scale structure of the universe but seems to fail at small scales \citep{Kravtsov10}. This is manifested by the departure of dark matter density profiles in centers of halos (at scales less than tens of kiloparsecs in clusters and even smaller in galactic halos) from the `cuspy' CDM profiles and the observed under-abundance of low-mass halos (with maximum circular velocities of less than a hundred kilometers per second) and the associated dwarf galaxies compared to the prediction of the CDM model. The weakly collisional two- (or multi-)component dark matter (2cDM) model --- named by analogy with Pontecorvo's `two-component neutrino' theory --- has a potential to resolve both problems. As in the self-interacting dark matter (SIDM) model \citep{SS00}, the central cusps are smeared out by collisions. However, unlike the SIDM model with a constant cross-section, the $\propto1/v$ dependence of the flavor-mixed cross-section in the 2cDM model reduces the core size in galaxy clusters (which have an order of magnitude larger velocity dispersion than galactic halos) and brings them in agreement with observations. In addition, in the very centers of halos, where the `optical depth' to collisions is much larger than unity, the dark matter will behave as a fluid and rapidly gravitationally collapse to form supermassive black holes \citep{HO02} thus providing a possible explanation for their existence at high redshifts. 

The observed paucity of small-mass halos can be due to the evaporation via $h\to l$ conversions discussed in this paper. [Note, SIDM and similar models cannot address this problem at all.] This evaporation occurs if the characteristic velocity, $v_k$, the mass eigenstates get upon conversion is comparable or exceeds the escape velocity of the halo. Cosmological simulations with two-component mixed dark matter, reported in detail elsewhere \citep{M13}, show that a number of CDM problems at small scales can be simultaneously resolved. For example, the break in the maximum circular velocity function of dark matter halos deduced from observation of dwarf galaxies to be at $v_c\sim50-100\text{ km s}^{-1}$ (corresponding to the halo mass of about $10^{10}$ solar masses) occurs when $v_c\sim v_k$, which implies the very high mass degeneracy $\Delta m/m\sim (v_k/c)^2\sim10^{-8}$. This is important for the following reason. The self-interaction cross-section should be rather large to be cosmologically interesting: $\sigma_\text{si}/m\gtrsim 0.1\text{ cm}^2\text{ g}^{-1}$, but cannot exceed $\sim{\cal O}(1)\text{ cm}^2\text{ g}^{-1}$ to not contradict observations. Such a value of cross-section corresponds to about one collision per the Hubble time (i.e., the age of the universe), which is still much longer than the decoherence time in a Milky Way-type, $10^{12}M_\odot$, halo. The decoherence time is even shorter for smaller ones and becomes comparable to the Hubble time only on scales of several megaparsecs. Thus, $m$-conversions and DM-evaporation can occur in all gravitationally bound systems. 

One would naively think that because of the large cross-section, all heavy eigenstates should be converted into the lightest one in the early universe soon after the dark matter freeze-out. This is not so, however, because the space-time is nearly flat in the early universe and the conversion cross-sections are strongly suppressed as compared to other interaction (e.g., scattering) cross-sections by a factor of $(\Delta m/m)^4\sim10^{-32}$, thus making conversion processes irrelevant at that time. The possibility of strong mass degeneracy is interesting for the direct detection experiments as well. In collisions of dark matter particles with matter in the detector, conversions of mass eigenstates can occur along with elastic scattering. Since mass-conversions change the kinetic energy of the particles by $\sim\Delta m\sim10^{-8}m$, such collisions will be slightly inelastic. The mass of dark matter particles is unknown, but if, for example, it is $m\sim\text{TeV}$, then the recoil energy detected in experiments can differ from the standard CDM prediction by $\sim\pm\Delta m\sim\pm10^{-8}m\sim\pm10~\text{keV}$ thus mimicking an inelastic dark matter. It would be very interesting to look for such a signal in the ongoing and future experiments.

\section{Conclusions}

In this paper we studied the evolution of non-relativistic interacting flavor-mixed particles. We demonstrated that particle-particle interactions can lead to inter-conversions of their mass eigenstates, in addition to elastic scattering. These conversions are most efficient when the mass eigenstates are well-separated in space, but they are suppressed in flat space-time (no gravity) for mass-degenerate eigenstates, Eq. (\ref{sigma-fs}). We stress that the conversions are not flavor oscillations: mass eigenstates remain intact during the oscillations. Also, conversions change momenta and kinetic energies of the eigenstates, Eqs. (\ref{pl-prime}), (\ref{deltav-deg}), which results in the effect called ``quantum evaporation''. Consider, for example, an ensemble of flavor-mixed particles trapped in a gravitational (or other non-flavor) potential well. Elastic collisions of the particles lead to the $m$-process, so the total number of particles trapped inside the well can decrease with time, Fig. \ref{Bmass}, because of the escape of the conversion ``secondaries''. Whether the evaporation will be complete or else some particles will remain trapped forever depends on the initial (flavor or mass) composition and the conversion cross-sections, Eq. (\ref{evap-condition}). We emphasize that unlike nuclear reactions, a particle kind does not change in conversions: neutrinos remain neutrinos, for example. What is changing is their flavor (and mass) composition, spatial localization, momenta and energies: some or all bound particles (i.e., with negative kinetic plus potential energy) can become free (with positive energy) without extra energy supplied to the system. Obviously, evaporation is different from tunneling: particle's energy does not change in the latter.  Finally, we discussed possible implications of the obtained results for (i) the cosmic neutrino background distortions in both the flavor composition and their energy and (ii) cosmology with two-component dark matter and possible resolutions of the core-cusp and substructure problems, and the better understanding of the early origin of supermassive black holes. A prediction for the direct detection dark matter experiments has also been made.

\begin{acknowledgments}
The authors is grateful to the Institute for Theory and Computation at Harvard University, where this works has been performed, for hospitality and the ITC colleagues for useful discussions. This work was supported in part by the ITC and by DOE grant  DE-FG02-07ER54940 and NSF grant AST-1209665. 
\end{acknowledgments}

\appendix
\section{Wave-packet dynamics and oscillations}
\label{A}

We assume, for illustration purpose, that a mixed particle is created in a definite flavor state $\alpha$ at $t=0$, $x_0=0$. The particle wave function is a superposition of mass eigenstates
\beq
\Psi_\alpha(x,0)=\Psi_h(x,0)\cos\theta + \Psi_l (x,0) \sin\theta.
\eeq
Hence
\beq
\Psi_h(x,0)=\Psi_\alpha(x,0) \cos\theta, \quad 
\Psi_l(x,0)=\Psi_\alpha(x,0) \sin\theta.
\eeq
At $t=0$, they can be chosen to be gaussian wave packets peaked at $x_0=0$, cf. Eq. (\ref{psi0}). The temporal evolution of these wave packets is
\beq
\Psi_h(x,t)=\psi_h(x,t) \cos\theta, \quad 
\Psi_l(x,t)=\psi_l(x,t) \sin\theta,
\eeq
where $\psi_j(x,t)$ with $j=h,l$ is given by Eq. (\ref{psi}). The thick red and blue curves in Fig. \ref{wavepackets} are the probability densities $|\Psi_h(x,t)|^2$ and $|\Psi_l(x,t)|^2$. The wave functions in the flavor basis are 
\bea
\Psi_\alpha(x,t)&=&\Psi_h(x,t) \cos\theta + \Psi_l(x,t) \sin\theta \nonumber\\
&=&\psi_h(x,t) \cos^2\theta + \psi_l(x,t) \sin^2\theta, \\
\Psi_\beta(x,t)&=& - \Psi_h(x,t) \sin\theta + \Psi_l(x,t) \cos\theta \nonumber\\
&=&-\left[\psi_h(x,t)  - \psi_l(x,t) \right]\cos\theta\sin\theta,
\eea
The thin cyan and magenta curves in Fig. \ref{wavepackets} are the probability distributions $|\Psi_\alpha(x,t)|^2$ and $|\Psi_\beta(x,t)|^2$. 

At this point, we make no assumptions about the initial wave packet widths ($\Delta_{0,h}$ may differ from $\Delta_{0,l}$), their energies and momenta. Moreover, the momenta and energy of the mass eigenstates are, in general, different and are determined by the process of production and/or detection (e.g., of the accompanying secondaries) \cite{AS09,Giunti01}. Solely for illustration purposes, we chose $\hbar=1,\ \theta=\pi/5,\ \Delta_{0,h}=\Delta_{0,l}=1,\ x_0=5,\ m_1=1.5,\ m_2=1,\ m_1 v_1=2,\ m_2 v_2=24,\ t=0.18$. 

The wave packet spreading is negligible at early times, so we can neglect it in the analytical analysis below. We have from Eq. (\ref{psi}):
\bea
\psi_j(x,t)&=& \sqrt{G_j(x,t)}
\exp\left[\frac{i}{\hbar}\left(m_jv_jx-\frac{m_jv_j^2}{2}t\right)\right],
\eea
where
\beq
G_j(x,t)=\frac{1}{\sqrt{2\pi}\Delta_{0,j}}\exp\left[-\frac{(x-v_j t)^2}{2\Delta_{0,j}^2}\right]
\eeq
is the gaussian shape function. The probability densities are calculated straightforwardly to yield
\bea
|\Psi_h(x,t)|^2&=&G_h(x,t)\cos^2\theta, \\
|\Psi_l(x,t)|^2&=&G_l(x,t)\sin^2\theta, \\
|\Psi_\alpha(x,t)|^2&=&G_h(x,t)\cos^4\theta + G_l(x,t)\sin^4\theta \nonumber \\
& & +2\sqrt{G_h(x,t)\,G_l(x,t)}\sin^2\theta\cos^2\theta \nonumber\\
& & \times\cos\left[\frac{m_l v_l}{\hbar}\left(x-\frac{v_l}{2} t\right)
-\frac{m_h v_h}{\hbar}\left(x-\frac{v_h}{2} t\right) \right], 
\nonumber\\ & &
\\
|\Psi_\beta(x,t)|^2&=&\cos^2\theta\sin^2\theta\,\biggl( G_h(x,t) + G_l(x,t) \nonumber \\
& & -2\sqrt{G_h(x,t)\,G_l(x,t)} \nonumber\\
& & \times\cos\left[\frac{m_l v_l}{\hbar}\left(x-\frac{v_l}{2} t\right)
-\frac{m_h v_h}{\hbar}\left(x-\frac{v_h}{2} t\right) \right]\biggr).
\nonumber\\ & &
\eea

\section{General analysis of cross-sections}
\label{B}

Suppose one of the mass eigenstates $\ket{m_i}$ (for example, $i=h$)  of a mixed particle is scattered off a potential $V({\bf r})$. Its initial wave function can be written as a plane wave, $\psi_i=e^{ik_i z}$, where $k_i =p_i /\hbar$ is the mass eigenstate wave number, $p_i$ is its momentum. The wave function in the elastic scattering ($i\to i$) channel is a superposition of the incoming and the elastically scattered waves:
\beq
\Psi_i=e^{ik_i z}+f_{ii}(\theta)\frac{e^{ik_i r}}{r}.
\eeq
In the inelastic (conversion, $i\to f$) channel, the wave function has the outgoing wave only:
\beq
\Psi_f=f_{fi}(\theta)\sqrt{\frac{m_f}{m_i}}\frac{e^{ik_f r}}{r}.
\eeq
The amplitudes of the elastic and inelastic processes, see e.g., \citep{LL}, are 
\bea
f_{ii} &=& \frac{1}{2ik_i}\sum_{l=0}^\infty(2l+1)\left(S_{ii}^{(l)}-1\right)P_l(\cos\theta), 
\\
f_{fi} &=& \frac{1}{2i\sqrt{k_i k_f}}\sum_{l=0}^\infty(2l+1)S_{fi}^{(l)}P_l(\cos\theta),
\eea
where $P_l(\xi)$ are the Legendre polynomials, $S_{ii}^{(l)}$ and $S_{fi}^{(l)}$ are the elements of the scattering matrix. These $S$-matrix elements are proportional to the $V$ matrix elements of Eq. (\ref{V}), that is $S_{ji}=\bra{m_j}V\ket{m_i}=\sum_{l=0}^\infty(2l+1)S_{ji}^{(l)}P_l(\cos\theta)$ with $j$ denoting the final state, which can coincide with the initial, as in elastic scattering ($j=i$), or be a different final state ($j=f$). These matrix elements are not completely independent. Unitarity requires that 
\beq
\sum_{j=i,f}\left|S_{ji}\right|^2 = \left|S_{ii}\right|^2 + \sum_{f}\left|S_{fi}\right|^2 = 1,
\eeq
where the latter sum is over all possible final states in inelastic channels. The Optical theorem follows directly from this equation. 

The differential (per solid angle) cross-sections of elastic scattering and of inelastic interactions ($m$-conversions) are
\bea
d\sigma_{ii} &=& \left|f_{ii}\right|^2\, d\Omega_i, 
\\
d\sigma_{fi} &=& \left|f_{fi}\right|^2\frac{p_f}{p_i}\, d\Omega_f.
\eea
The total cross-sections are
\bea
\sigma_{ii} &=& \frac{\pi}{k_i^2}\sum_{l=0}^\infty(2l+1)\left|1-S_{ii}^{(l)}\right|^2,
\label{sigma-ii}\\
\sigma_{fi} &=& \frac{\pi}{k_i^2}\sum_{l=0}^\infty(2l+1)\left|S_{fi}^{(l)}\right|^2.
\label{sigma-fi}
\eea
Note that is there is elastic scattering only, then $S_{fi}^{(l)}=0$ and $|S_{ii}^{(l)}|=1$; in the presence of inelastic channels ($m$-conversions) $|S_{ii}^{(l)}|<1$. When the elastic scattering matrix element vanishes, $S_{ii}^{(l)}=0$ for a certain $l$, the initial particle with angular momentum $l$ is completely `absorbed', so only $m$-conversions have non-zero amplitudes. Then the partial cross-sections for elastic and all conversions are equal 
\beq
\sigma_{ii}^{(l)}=\sum\nolimits'_f\sigma_{fi}^{(l)}=(\pi/k_i^2)(2l+1),
\label{sigmas}
\eeq 
where $\sum'_f$ is the sum over all $f\not=i$.

Let the scattering potential have a characteristic radius of interaction, $r_0$. For slow particles, whose wavelength is much larger than the size of the potential, $r_0 p/\hbar=r_0k\ll1$, the partial amplitudes with large angular momentum $l$ decay so that $\sigma^{(l)}_{ii}\propto k^{4l}$ and $\sigma_{fi}^{(l)}\propto k^{2l-1}$ if elastic scattering dominates over inelastic processes \citep{LL}, so the $l=0$ term dominates ($s$-wave scattering) and other terms can be safely neglected:
\bea
f_{ii} &\simeq& \frac{\left(S_{ii}-1\right)}{2ik_i}, 
\label{amplitudes1}
\\
f_{fi} &\simeq& \frac{S_{fi}}{2i\sqrt{k_i k_f}}.
\label{amplitudes}
\eea
This is relevant to neutrino systems interacting weakly and, possibly, to dark matter particles, at least if the also interact via weak forces. Hereafter the superscript $(l)$ is omitted for clarity. 

Let us consider two limiting cases. First, if elastic scattering dominates over conversions, $\left|S_{ii}\right|^2\sim1$, then as usual $S_{ii}-1=e^{2i\delta_0}-1\simeq 2i\delta_0= 2ik_i\alpha$, where $\delta_0$ is the phase induced by scattering and $\alpha=\alpha'-i\alpha''$ is the complex scattering length and $\alpha',\ \alpha''$ are real constants, which is so for a sufficiently spatially localized scattering potential. Then we recover the standard result
\bea
& & {}\sigma_{ii} = 4\pi |\alpha|^2, 
\label{sigma-el-1}\\
& & {}\sum\nolimits'_f\sigma_{fi}=(4\pi/k_i)|\alpha''|^2, 
\label{sigma-con-1}
\eea
that is the elastic cross-section is independent of particle's velocity, $\sigma_{ii}\propto const.$, and the conversion cross-sections scale as $\sigma_{fi}\propto 1/v$.

Second, alternatively, when the $i\to i$ S-matrix element vanishes, $S_{ii}=0$, the cross-sections, from Eqs. (\ref{sigmas})-(\ref{amplitudes}), are
\bea
\sigma_{ii} &=& \sum\nolimits'_f \sigma_{fi} = \pi /k_i^2, 
\label{sigma-el-2}\\
\sigma_{fi} &=& (\pi/k_i^2) \left| S_{fi}\right|^2,
\label{sigma-con-2}
\eea
that is they scale as $\sigma\propto 1/v^2$. Note that $\sum\nolimits'_f \left| S_{fi}\right|^2=1$ due to unitarity. We emphasize that there is no divergence in summing over all $l$. Indeed, the above expressions describe an interacting particle with the angular momenta $l<kr_0=l_0$ or equivalently, with the impact parameter $\rho=l/(mv/\hbar)<r_0$. Thus, the sum is finite: the terms with $0\le l \la l_0$, and the higher-$l$ terms decay fast for a well localized potential.

Now we outline how this analysis extends to a system of two interacting particles. In this case, both the initial state and the final state consist of two mass eigenstates located far apart. We can choose one of them to be a scatterer (target, $t$) and consider how another (a projectile, $s$) is scattered off it, $s_i t_i\to s_f t_f$, where the initial and final $s_{i,f}$ and $t_{i,f}$ can be any combination of $h$ and $l$ (the amplitudes of kinematically forbidden processes vanish). If the initial mass eigenstates are different, as is in the case of $lh\to...$ and $hl\to...$ processes, than the projectile wave function is simply
\beq 
e^{i k_{i} z},
\eeq
where $i$ is $h$ or $l$. However, if they are identical, which corresponds to $hh\to...$ and $ll\to...$ processes, one cannot tell apart the target and the projectile. Instead, the wave function is even or odd with respect to the particle interchange, hence
\beq 
e^{i k_i z} \pm e^{i k_i z}, 
\eeq
depending on whether the total spin of the particles is even or odd, respectively. For instance, for two half-spin particles, the former corresponds to the zero total spin and the latter to the total spin equal to one.  The above wave function represents one particle coming from the left and another from the right. 

The outgoing (scattered) wave function is also written differently, depending on whether the outgoing mass eigenstates are different or the same. In the former case,  i.e., for the processes like $\dots\to lh$ and $\dots\to hl$, the outgoing particle wave  function is simply 
\beq
f_{fi}(\theta)\sqrt{\frac{m_f}{m_i}}\frac{e^{ik_f r}}{r}.
\eeq
In the latter case, i.e., for the processes like $\dots\to ll$ and $\dots\to hh$, the outgoing particles are identical (i.e., $f_1=f_2=f$) so the particle wave function may be either even or odd with respect to the particle interchange
\beq
\left[f_{fi}(\theta)\pm f_{fi}(\pi-\theta)\right]\sqrt{\frac{m_f}{m_i}}\frac{e^{ik_f r}}{r},
\eeq
which again depends on whether the total spin is even or odd. 

Obviously, the cross-sections, being proportional to $\left|f_{fi}(\theta)\pm f_{fi}(\pi-\theta)\right|^2$,  will have the interference terms $f_{fi}(\theta) f^*_{fi}(\pi-\theta)+f^*_{fi}(\theta) f_{fi}(\pi-\theta)$. Whether they are important or now depends on the shape of the interaction potential and the particles' relative velocity. For example, the total elastic cross-section of Coulomb scattering approaches the classical Rutherford cross-section only in the limit of small velocities $v\ll e^2/\hbar$. 
In general, the total elastic cross-section is 
\beq
\sigma=\frac{s}{2s+1}\sigma_+ + \frac{s+1}{2s+1}\sigma_- ,
\eeq
where the symmetric and antisymmetric cross-sections are
\beq
\sigma_\pm = \int\left|f_{fi}(\theta)\pm f_{fi}(\pi-\theta)\right|^2\, d\Omega
\eeq
and $d\Omega$ is the solid angle element. We note that when $s$ wave scattering dominates (e.g., for slow particles), the scattering amplitudes tend to a constant, hence the antisymmetric  cross-section vanishes. Thus, only particles with the even total spin can scatter. Further analysis of such details goes beyond the scope of this paper.

\end{document}